# Conic Sections in Ferroelectric Nematics: Experiments and Mathematical Modeling

Priyanka Kumari[1,2], Olexandr Kurochkin[3,4], Vassili G. Nazarenko[3,4], Oleg D. Lavrentovich[1,2,5]*, Dmitry Golovaty[6], and Peter Sternberg[7]

[1]*Advanced Materials and Liquid Crystal Institute, Kent State University, Kent, OH 44242, USA*
[2]*Materials Science Graduate Program, Kent State University, Kent, OH 44242, USA*
[3]*Institute of Physics, National Academy of Sciences of Ukraine, Prospect Nauky 46, Kyiv, 03028, Ukraine*
[4] *Institute of Physical Chemistry, Polish Academy of Science, Kasprzaka 44/52, 01-224 Warsaw, Poland*
[5]*Department of Physics, Kent State University, Kent, OH 44242, USA*
[6] *Department of Mathematics, The University of Akron, Akron, OH 44325, USA*
[7]*Department of Mathematics, Indiana University, Bloomington, IN 47405, USA*
*Corresponding author, olavrent@kent.edu*

**Abstract:** Domain structure of a fluid ferroelectric nematic is dramatically different from the domain structure of solid ferroelectrics since it is not restricted by rectilinear crystallographic axes and planar surface facets. We demonstrate that thin films of a ferroelectric nematic seeded by colloidal inclusions produce domain walls in the shape of conics such as a parabola. These conics reduce the bound charge within the domains and at the domain walls. An adequate description of the domain structures requires one to analyze the electrostatic energy, which is a challenging task. Instead, we demonstrate that a good approximation to the experimentally observed polydomain textures is obtained when the divergence of spontaneous polarization - which causes the bound charge is heavily penalized by assuming that the elastic constant of splay in the Oseen-Frank energy is much larger than those for twist and bend. The model takes advantage of the fact that the polarization vector is essentially parallel to the nematic director throughout the sample.

## 1. Introduction

Solid ferroelectrics are polydomain. Within each domain, the spontaneous electric polarization **P** aligns along a certain rectilinear crystallographic axis (1-4). As first proposed by Landau and Lifshitz (5), the domains form in response to a finite size of samples in order to reduce depolarization fields. Domains with a differently oriented **P** are separated by domain walls (DWs), which are generally flat, as dictated by crystallographic axes and crystal facets (1-6).

The recently discovered ferroelectric nematic liquid crystal (N$_F$) (7-10) is a liquid with a macroscopic spontaneous polarization **P**. This polarization is locally parallel to the director $\hat{\mathbf{n}} \equiv -\hat{\mathbf{n}}$, which specifies the average quadrupolar molecular orientation (11). The polarization direction could be aligned by confining the material between two glass plates with rubbed polymer coatings (9, 10, 12-20). In these samples, the DW shape is defined by the anisotropic surface interactions with the "easy axis" of the substrate and by the orientational elasticity of N$_F$. The DWs in a surface-aligned N$_F$ are rectilinear (13, 19), zig-zag (12, 16-18), lens-like (10, 18), or smoothly curved (9, 10, 12, 15, 17, 18, 20). In samples with air bubbles, one observes incomplete parabolic walls which separate concentric patterns of the polarization imposed by the air-nematic interface and a uniform domain set by a rectilinear easy axis at the substrate (21). Experiments with fully degenerate in-plane surface anchoring (22) reveal that the prevailing type of domains not constrained by crystallographic axes and azimuthal anchoring are domains with (a) nearly uniform polarization or (b) nearly circular polarization, which implies bend deformation of **P**. Splay of **P** is diminished because it creates a bound ("space") charge of bulk density $\rho_b = -\text{div}\,\mathbf{P}$ and thus increases the electrostatic energy. The avoidance of polarization splay in defect textures has been described previously for ferroelectric smectics C by Link et al. (23). In the azimuthally degenerate N$_F$ films, DWs separating a circular and a uniform domain are parabolas (eccentricity $e = 1$), while DWs between two circular domains are hyperbolas ($e > 1$) (22). The eccentricity can vary along the DW; as a rule, $e < 1$ near the tip of the DW. The observed domain textures minimize the bound electric charge of bulk density $\rho_b = -\text{div}\,\mathbf{P}$ and of the surface density $\sigma_b = (\mathbf{P}_1 - \mathbf{P}_2) \cdot \hat{\mathbf{k}}$ at the DWs separating two neighboring polarization patterns $\mathbf{P}_1$ and $\mathbf{P}_2$. Here $\hat{\mathbf{k}}$ is the unit normal to a DW, pointing towards domain 1 (22, 24). To reduce $\sigma_b$, a DW must bisect the angle between $\mathbf{P}_1$ and $\mathbf{P}_2$, so that $\mathbf{P}_1 \cdot \hat{\mathbf{k}} = \mathbf{P}_2 \cdot \hat{\mathbf{k}}$, making the projection of polarization onto $\hat{\mathbf{k}}$ continuous across the DW, while the tangential component changes sign. The remarkable bisecting properties of conics have been elucidated millennia ago by Apollonius of Perga (25). However, the bound charges are still present when the polarization realigns continuously in the plane of the sample along $\hat{\mathbf{k}}$ over a finite DW width. The projection $\mathbf{P} \cdot \hat{\mathbf{k}} = P_k$ of the polarization onto $\hat{\mathbf{k}}$ yields a non-vanishing bound charge density $-\partial P_k/\partial k$. This produces two oppositely charged sheets at the DW (21, 26, 27). The situation is reminiscent of electrically charged Néel walls in solid ferroelectrics, in which the polarization realigns in a plane perpendicular to the wall (28). The presence of some bound charge in the $N_F$ samples is also evidenced by observations of $2\pi$ soliton DWs with splay-bend realignment of $\mathbf{P}$ being topologically protected (19) and of DWs separating circular domains with opposite sense of polarization circulation (22).

A full description of polarization patterns with the bound charges should involve the analysis of the (partially screened by ions) electrostatic energy (29), which is a difficult task. One often uses a simplified model, in which the electrostatics is reduced to a renormalization of the splay elastic constant (30-33), $K_1 = K_{1,0}(1 + \lambda_D^2/\xi_P^2)$. Here $K_{1,0} \sim 10$ pN is the bare splay modulus of the same order as the one measured in a conventional paraelectric nematic (N), $\lambda_D = \sqrt{\frac{\varepsilon\varepsilon_0 k_B T}{n e^2}}$ is the Debye screening length, $\xi_P = \sqrt{\frac{\varepsilon\varepsilon_0 K_{1,0}}{P^2}}$ is the polarization penetration length, $\varepsilon_0$ is the electric constant, $\varepsilon$ is the dielectric permittivity of the material, $e = 1.6 \times 10^{-19}$ C is the elementary charge, $n$ is the concentration of ions, $k_B$ is the Boltzmann constant, and $T$ is the absolute temperature. For the typical $n = 10^{23}$ m$^{-3}$, $\varepsilon = 10 - 100$ (34, 35), $P = 6 \times 10^{-2}$ C/m$^2$ (10), and at room temperature, one finds $\lambda_D \approx (10 - 30)$ nm and $\xi_P \approx (1 - 2)$ nm. Note that here we do not use the often-reported exaggerated values of $\varepsilon$ since these represent an artifact of dielectric measurements in $N_F$ cells (34, 36). Since $\lambda_D > \xi_P$ (10), $K_1$ in the $N_F$ should be much larger than $K_{1,0}$ in the N and larger than the twist $K_2$ and bend $K_3$ elastic constants in the $N_F$.

Experimental data on elastic properties of $N_F$-forming materials are scarce. Chen et al. (18) measured $K_1 \approx 10 K_2$ in the N phase of DIO and expected $K_1 \approx 2$ pN (33). Mertelj et. al.

(37) reported that in the N phase of the ferroelectric material RM734, $K_1$ is even lower, about 0.4 pN. Since the bend elastic constant $K_3$ of N$_F$ does not experience an electrostatic renormalization, it is expected to be a few tens of pN; Mertelj et. al. (37) found $K_3 \approx$ 10-20 pN for the N phase of RM734. Therefore, the ratio $K_1/K_3$ in the N$_F$ could be larger than 1, ranging from single-digits to $\sim 10^2$. Studies of $2\pi$ DWs in N$_F$ (19) suggest $K_1/K_3 > 4$.

The goal of this work is to explore whether a model with a strong disparity of the elastic constants, $K_1 \gg K_2, K_3$, can explain the experimentally observed polarization patterns with DWs in the shape of conic sections. In Section 2, we present the experimental optical microscopy textures of these patterns. In Section 3, we propose a model which employs the well-known director-based Oseen-Frank energy with splay, bend, twist, and saddle-splay terms

$$E_{OF}(\hat{\mathbf{n}}) := \int_{\Omega} [\frac{K_1}{2}(\text{div }\hat{\mathbf{n}})^2 + \frac{K_2}{2}(\hat{\mathbf{n}} \cdot \text{curl }\hat{\mathbf{n}})^2 + \frac{K_3}{2}|\hat{\mathbf{n}} \times \text{curl }\hat{\mathbf{n}}|^2$$
$$+ \frac{K_2 + K_4}{2}(\text{tr }(\nabla \hat{\mathbf{n}}^2) - (\text{div }\hat{\mathbf{n}})^2)] \, dx_1 \, dx_2 \, dx_3,$$

for a nematic liquid crystal occupying a region $\Omega$. However, since in the materials under consideration, the polarization vector tends to align with the director $\hat{\mathbf{n}}$, we replace $\hat{\mathbf{n}}$ with $\mathbf{P}$ when modeling these ferroelectric nematic textures, and ignore associated orientability issues that may arise. This Oseen-Frank type energy for the polarization is supplemented by a potential term that serves to set the preferred value for the magnitude of polarization, and by an anchoring term that strongly favors tangential anchoring on the surface of the sample, thus penalizing the component of $\mathbf{P}$ normal to the surface. Most crucially, we pursue an asymptotic regime where the cost of splay is dominant over the other terms in the elastic energy density, that is $K_1 \gg K_2, K_3$. To simplify the model further, we ignore saddle splay by setting $K_2 + K_4 = 0$.

## 2. Experiments

We explore two N$_F$ materials, abbreviated DIO (8) and RM734 (7). On cooling from the isotropic (I) phase, the phase sequence of DIO, synthesized as described previously (19), is I $-$ 174°C $-$ N $-$ 82°C $-$ SmZ$_A$ $-$ 66°C $-$ N$_F$ $-$ 34°C $-$ Crystal, where SmZ$_A$ is an antiferroelectric smectic (18). RM734 of purity better than 99% is purchased from Instec, Inc. The material is additionally purified by silica gel chromatography and recrystallization in ethanol. Its phase sequence is I—188°C —N—133°C —N$_F$—84°C —Crystal. The N$_F$ samples are of three types:

(i) DIO films with degenerate azimuthal surface anchoring are prepared by depositing a small amount of DIO onto the surface of glycerol (Fisher Scientific, CAS No. 56-81-5, assay percent range 99-100% w/v) with density 1.26 g/cm$^3$ at 20 °C in an open Petri dish. A piece of crystallized DIO is placed onto the surface of glycerol at room temperature, heated to 120 °C, then cooled down to the N$_F$ phase at a rate of 5 °C/min. DIO spreads over the surface and forms a film of an average thickness $h$ defined by the known deposited mass $M$ and the measured area $A$ of film, $h = M/\rho A$, where $\rho$ = (1.32–1.36) g/cm$^3$ is the density of DIO (5). The film shows two types of domains: domains of nearly uniform polarization and domains with circular polarization; both tend to avoid splay of **P**, Fig.1(a-c). To gain a better control on the occurrence of circular domains and to facilitate comparison with the numerical simulations, in which the circular domains should be introduced artificially, say, by boundary conditions, in some samples we seed circular vortices of polarization by a small number of colloidal SiO$_2$ silica spheres. The spheres of a diameter 6 μm are added to the liquid crystal in the N phase; upon cooling to the N$_F$ phase, each sphere induces a circular domain of **P**, Fig.1(d-g).

(ii) Flat cells of RM734 are assembled from glass plates spin-coated with thin (50 nm) layers of polystyrene, separated by a distance $h$ = (1–10) μm and sealed with an epoxy glue Norland Optical Adhesive (NOA) 65. Polystyrene aligns **P** tangentially (22, 38). Silica spheres are added to some samples, Fig.2.

(iii) Freely suspended DIO films formed in square openings of metallic grids used as holders for samples in transmission electron microscopy. A 5wt% solution of DIO in dodecane is heated to 90 °C and then spread across the openings. After solvent evaporation, freely suspended DIO films form.

Below we analyze the shape of the DWs and the type of deformations they carry.

### 2.1. Bend and splay of polarization in DWs.

To analyze the textures, we use both the conventional polarizing optical microscopy and PolScope approach, invented by Oldenburg (39, 40), applications of which to liquid crystals has been described in Refs. (41, 42). Briefly, a PolScope represents a polarizing optical microscope with a variable optical compensator(s), which might be a nematic liquid crystal cell controlled by an electric field. The image of a sample is recorded in polarized light multiple times with different settings of the optical compensator; the numerical analysis of the set of transmitted light intensity maps reconstructs the two-dimensional maps of the optic axis (in projection onto the plane of imaging) and optical retardance. The approach assumes that the optic axis does not change along the light propagation direction. The PolScope observations in this study are performed by the Exicor Microimager (Hinds Instruments) operating at four wavelengths, 475 nm, 535 nm, 615 nm, 655 nm, which allows one to characterize samples with optical retardance up to 3500 nm.

The textures in samples (i) and (ii) show that the colloidal spheres trigger circular domains of polarization and DWs of parabolic shape separating such a circular domain from a domain with a nearly uniform polarization, Fig.1. In particular, Figure 1(a) shows the in-plane map of the optic axis, which is parallel to the director and to $\mathbf{P}$ in the studied materials. The DW shapes are fitted with an equation of a conic, written in polar coordinates $(r, \psi)$ with the origin at the core of a circular vortex,

$$\frac{d}{r} = \frac{1}{e} - \cos\psi, \tag{1}$$

where $e$ is the eccentricity, and $d$ is the distance from the core to the directrix. The fitted values are listed in Figs.1, 2, and 3; the accuracy is better than 5%. The eccentricity of DWs separating a circular vortex and a uniform domain is close to 1, hence the name "P-wall", where "P" stands for the "parabolic" (22). The tip region is often an exception since there $e$ can be much smaller than 1; for example, $e = 0.12$ at tip of the DW in Fig.1(f). This region is called a "T-wall" (22) to stress that the polarizations $\mathbf{P}_1$ and $\mathbf{P}_2$ on opposite sides of the wall are tangential to it (and antiparallel to each other), Fig.1(a). In the samples with colloidal seeds, the eccentricities deviate from 1 rather strongly, by $\pm 0.25$, Fig.1(d-g) and Fig.2. An apparent reason is the meniscus around the colloidal spheres, which implies a nonzero dihedral angle between the surface of the sphere and the $N_F$ interfaces with air and glycerol in Fig.1(d-g) and polystyrene-coated glass plates in Fig.2. The resulting thickness gradients create a torque forcing $\mathbf{P}$ to be perpendicular to the gradient direction, in order to avoid splay (43).

In solid ferroelectrics, the DW are often of the Ising type. For example, in a $\pi$-wall of the Ising type separating two domains with antiparallel polarizations, the polarization remains parallel to the same crystallographic axis, but its magnitude decreases to zero, $|\mathbf{P}| \to 0$, in the middle of the wall (28). The parabolic DWs in our experiments are different from the Ising DWs as they do not show any significant decrease of the polarization magnitude, as revealed by strong birefringence observed within the entire width $w \sim 10$ μm of the DW, Fig.1(a). The optical retardance across the wall changes continuously and smoothly, Fig.1(a), lacking the abrupt discontinuity expected in an Ising wall. Beside strong birefringence, another argument against a "polarization melting" within the wall is a high energy of such a melting, which can be estimated, following a similar approach proposed by de Gennes for the nematic-to-isotropic transition (11), as $k_B T/V \approx 6 \times 10^6$ J/m$^3$, where $k_B = 1.38 \times 10^{-23}$ J/K is the Boltzmann constant, $T \approx 400$ K is the approximate temperature of the phase transition from the N$_F$ to an antiferroelectric or paraelectric phase, and $V \approx 1$ nm$^3$ is the molecular volume. This energy density is much higher than the elastic energy density of reorientation of $\mathbf{P}$ within the DW, estimated as $\frac{\overline{K}}{w^2} \sim 10$ J/m$^3$, where $\overline{K}$ is some average of the bend and splay elastic constants, taken for the purpose of this comparison to be of an exaggerated value $10^3$ pN. Instead of complete polarization melting, the prevailing mode of connection of two neighboring polarization domains $\mathbf{P}_1$ and $\mathbf{P}_2$ is through realignment of the polarization, as in the Bloch and Néel walls of solid ferromagnets; some variation of the absolute value of $\mathbf{P}$ should not be excluded.

The overall parabolic shape of the P-wall separating a domain with a uniform polarization $\mathbf{P}_1$ and a domain with a circular polarization $\mathbf{P}_2$ is explained by the avoidance of the bound charge on it, i.e., $\sigma_b = (\mathbf{P}_1 - \mathbf{P}_2) \cdot \hat{\mathbf{k}} = 0$, Fig.3. The last condition is fulfilled when the angle $\theta_1$ between $\mathbf{P}_1$ and the DW is equal to the angle $\theta_2$ between $\mathbf{P}_2$ and the DW, $\theta_1 = \theta_2 = \theta$, Fig.3(a). These two angles increase as one approaches the vertex of the parabola, located at the origin $(x, y) = (0,0)$ of the Cartesian coordinates: $\theta_1 = \theta_2 = \theta = \arctan\sqrt{x/f}$, where $f$ is the distance between the vertex and the focus, i.e., the center $(f, 0)$ of the circular domain, Fig.3a. However, instead of the cusp-like singularity at the merging $\mathbf{P}_1$ and $\mathbf{P}_2$, the polarization vector within the DW realigns smoothly from $\mathbf{P}_1$ to $\mathbf{P}_2$ over a finite width $w \approx$ (10-30) μm of the wall, Fig.1(a).

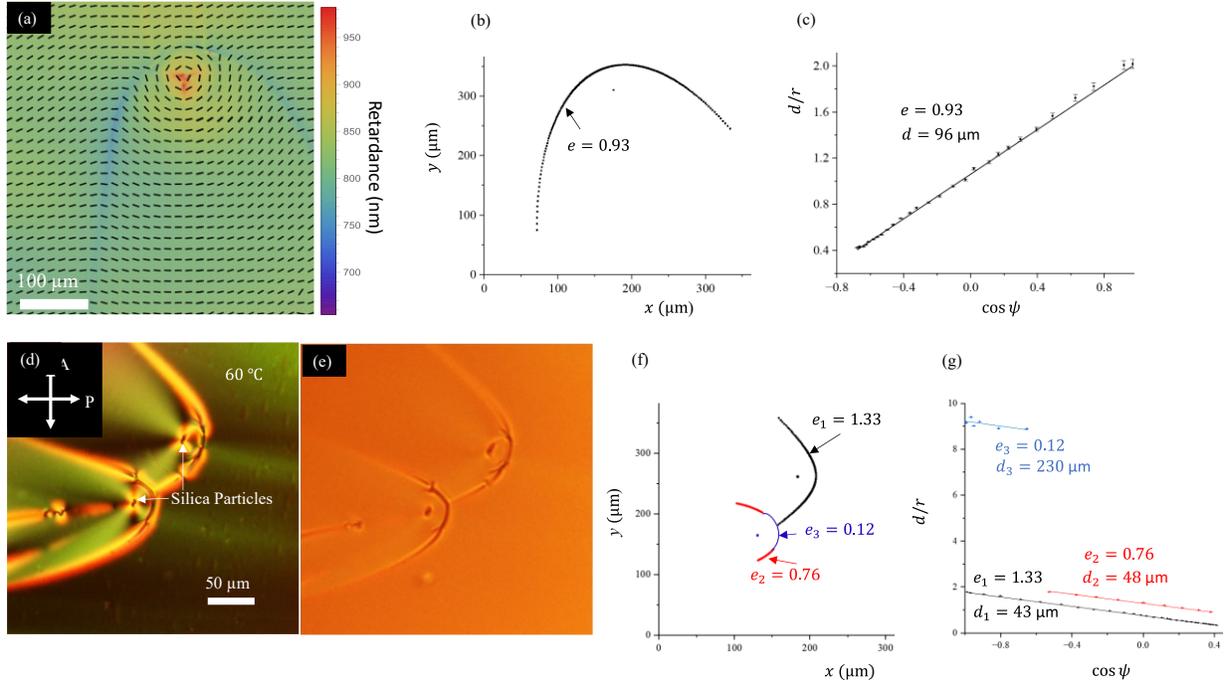

**Fig.1.** DWs in DIO $N_F$ films at glycerol. (a) PolScope Microimager texture of a parabolic DW with two -1/2 disclinations near the tip. The ticks show the local orientation of the optic axis which is parallel to **P**. The high retardance at the core of the circular domain is an artefact caused by the crossover into a different interference order. DIO film of thickness $h = 4.3$ μm. (b,c) Parabolic DW shape of eccentricity $e = 0.93$ fitted by Eq.(1). (d) Polarizing optical microscopy texture of an $N_F$ film, $h \approx 6$ μm; crossed polarizers. (e) The same, no analyzer. (f,g) DW shapes fitted by Eq.(1).

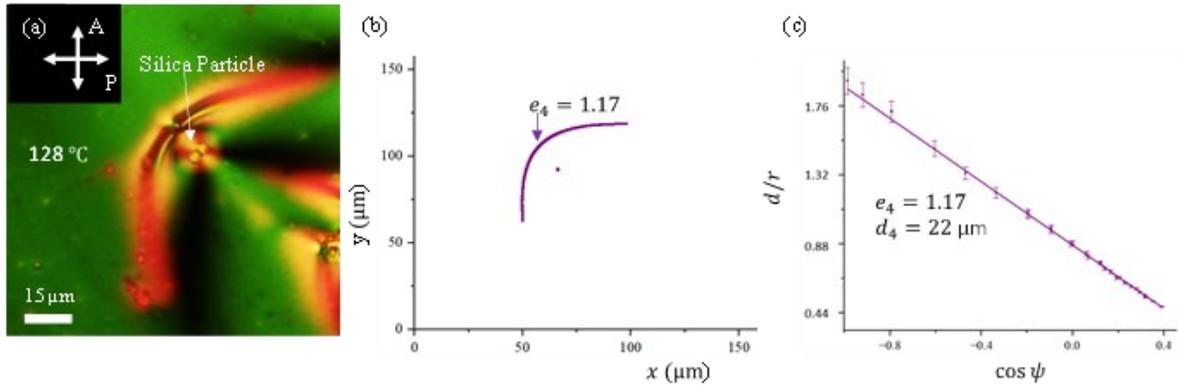

**Fig.2.** DWs in RM734 $N_F$ cells bounded by two polystyrene coated glass plates. (a) Polarizing optical microscopy texture; crossed polarizers; $h = 7$ μm. (b,c) DW shape fitted by Eq.(1).

The inner structure of the parabolic DW is different far away from the vertex, where $e \approx 1$ and near it. Far away from the vertex, **P** realigns from $\mathbf{P}_2$ to $\mathbf{P}_1$ by a small angle $\delta = \pi - 2\theta$, Figs.1(a), 3. As is easy to see, $\delta(x \to \infty) \to 0$ and the associated elastic energy is low. As one approaches the vertex, the bend angle increases, $\delta(x \to 0) \to \pi$, and the polarization realignment resembles a hairpin of a high elastic energy, Fig.3(a), thus necessitating restructuring of the DW. A common type of restructuring is the appearance of two -1/2 disclinations separating the parabolic DW branches with $e \approx 1$ from the tip of the DW, Fig.3(a). The -1/2 disclinations are clearly seen in PolScope Microimager texture in Fig.1(a). The tip region sandwiched between the two -1/2 disclinations is the T-wall. It is much more narrow (a few micrometers) and of a sharper optical contrast than the P-branch of the DW, Fig.1(d,e), which allows one to distinguish the two segments of DW even if the entire DW can be fitted by a conic of a constant eccentricity, Fig.1(b,c). Within the T-branch, the polarizations $\mathbf{P}_1$ and $\mathbf{P}_2$ tend to be antiparallel to each other and parallel to the wall, while at the P-branch of the same DW, the polarization *crosses* the wall while realigning from $\mathbf{P}_1$ to $\mathbf{P}_2$ that are not collinear. This realignment involves energetically costly splay, which might be the reason why the P-branch is wider than the T-branch. Sometimes, the -1/2 disclinations coalesce so that the T-wall degenerates into a -1 disclination (22). Similarly, the +1 core of the circular domain sometimes splits into a pair of +1/2 defects, for example, as a result of shear in the film. Note that the combination of one +1 defect at the center of a circular domain and two -1/2 disclinations (or two -1/2 and two +1/2, etc.) produces a zero topological charge of the entire structure; the far field of polarization is topologically trivial. Each -1/2 disclination replaces a hairpin with a large bent angle $\delta = \pi - 2\theta$ of polarization along the P-branch with a much smaller misalignment of $\mathbf{P}_1$ and $\mathbf{P}_2$, $2\beta \approx 2\theta(x \to 0) \to 0$ along the T-branch of the same DW, Fig.3(a). Since on one side of a T-wall there is a domain of a circular polarization $\mathbf{P}_2$, it explains why its shape is elliptical or nearly circular, with $e \ll 1$. The nearly circular T-wall creates a bend in the domain of a nearly uniform polarization $\mathbf{P}_1$, Fig.1(a). This bend produces a "ghost" parabola, which extends outside the prime parabolic DW and often originates from the cores of the -1/2 disclinations, Fig.4.

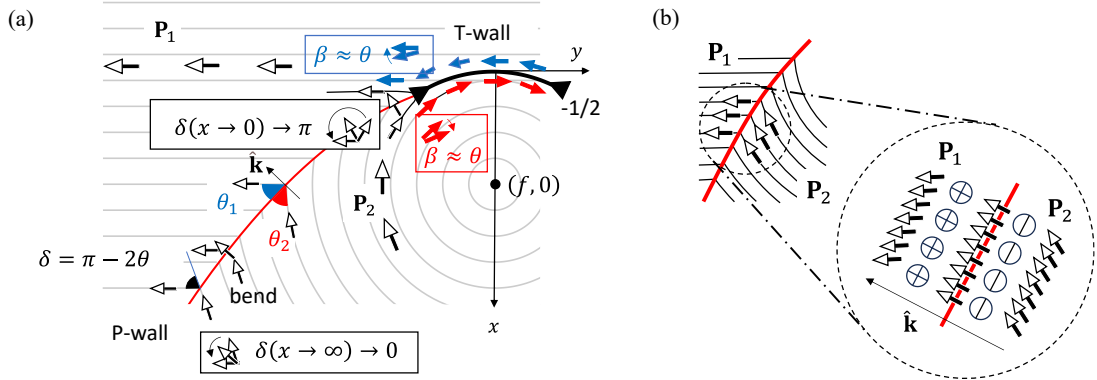

**Fig.3.** Schematic structure of a parabolic DW. (a) Polarization realignment from the uniform domain field $\mathbf{P_1}$ to the circular domain $\mathbf{P_2}$ across the DW. At the tip of parabola, high elastic energy is relieved by two -1/2 disclinations (black triangles) and a T-wall joining them; each disclination replaces one large mismatch angle $\delta = \pi - 2\theta$ with two small angles $\beta \approx \theta$. The nearly circular T-wall bends the outside polarization $\mathbf{P_1}$; compare to Fig.1(a). (b) The electric double layer around the DW caused by a nonzero $-\partial P_k/\partial k$; the "+" and "−" symbols inside circles refer to the bound charge and should not be confused with the free charge of ions.

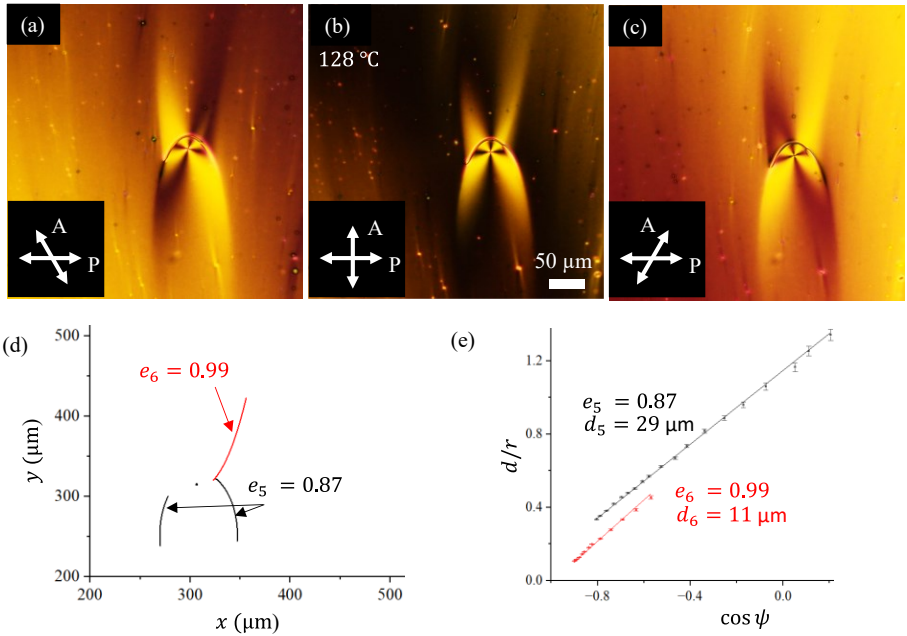

**Fig.4.** A parabolic DW with a T-wall at the tip creating a ghost parabolic DW in RM734 $N_F$ confined between two polystyrene coated glass plates. (a-c) Polarizing optical microscope textures with the polarizer and analyzer making an angle 120°, 90°, and 60°, respectively. The cell thickness $h \approx 4.3$ μm. (d) The fitted shapes of the DWs. (e) The corresponding eccentricities $e$ and distance $d$ to directrices fitted with Eq. (1).

## 2.2. Twists of polarization in DWs.

Solid ferroelectrics exhibit Néel DWs, in which the polarization experiences splay-bend realignment in the plane perpendicular to the wall, and Bloch DWs, in which the polarization twists along the axis perpendicular to the wall (28). The parabolic DWs in our study reveal splay-bend deformations of **P** in the plane of the sample, as discussed in the previous section, Figs.1-3, but also twists.

The twists along the normal to the sample, i.e., along an axis in the DW plane, are easy to uncover under a polarizing optical microscope with decrossed polarizers, Fig.5. In Fig.5, segments 1,3,5 of the parabolic DW and segments 2,4 alternate in brightness when the analyzer is rotated clockwise, Fig.5a, and counterclockwise, Fig.5c, with respect to the polarizer. The observations suggest that **P** in the segments 1,3,5 experiences a right-handed twist around the $z$-axis normal to the film, while segments 2,4 exhibit a left-handed twist. Since the twist $z$-axis is in the plane of the wall, the parabolic DW is different from the twist Bloch walls in solid ferroelectrics, in which the twist axis is perpendicular to the wall. The regions with opposite twists are separated by line defects, Fig.5.

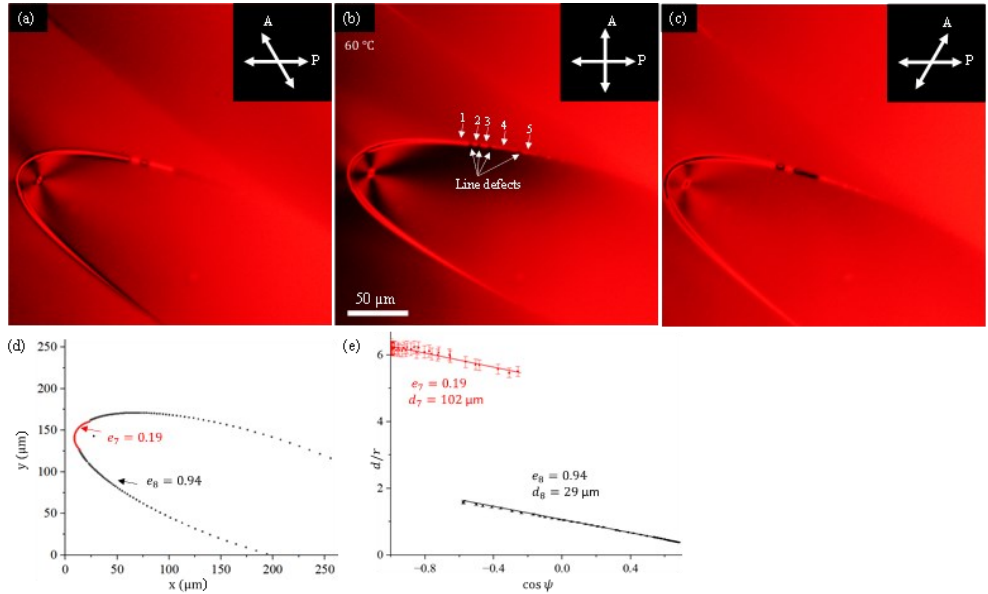

**Fig.5.** Alternating twists within the parabolic DW DIO film on glycerol, monochromatic light. (a-c) Polarizing optical microscope textures with the polarizer and analyzer making an angle 120°, 90°, and 60° respectively. Film thickness $h = 7\ \mu m$. (d) The fitted shapes of the DWs. (e) The corresponding eccentricities $e$ and distance $d$ to directrices as fits to Eq. (1).

Alternating left and right-handed twists of $\mathbf{P}$ around the $z$-axis normal to the film exist even in the geometry when a uniform unidirectional alignment, say, $\mathbf{P} = (0, P, 0)$ corresponds to the Oseen-Frank *elastic* (but not the *electrostatic*) energy minimum. These twisted states are observed in films with one surface providing a unidirectional alignment and the opposite surface being azimuthally degenerate (as the surfaces in the present study) (44). Although the twists increase the elastic energy, they reduce the electrostatic energy (44).

Twists are also apparent in the T-walls, as these walls do not show complete extinction in the segments that are parallel to the polarizer or analyzer. This feature also suggests that the polarization does not form a homeotropic region $\mathbf{P} = (0, 0, P)$ in the center of the wall, as would be the case for a pure Bloch wall. It is likely that the $z$-component of polarization varies with $z$ within the DW and thus introduces twist along a horizontal axis accompanied by splay-bend near the interfaces, as described for $2\pi$ walls previously (19). The twist with a non-zero $z$-component reduces the elastic energy of splay necessitated by the in-plane $\pi$ −realignments of $\mathbf{P}$ (19). The small width of the T-walls makes it difficult to decipher the complex deformation field solely by optical microscopy. Nevertheless, it is safe to conclude that the T- and P-walls represent a complex mix of all three bulk deformations, splay, bend, and twist.

### 2.3. Walls or surface disclinations.

Twist deformations are a common feature of other DWs in the $N_F$, as they have been previously identified in $2\pi$ (19) and $\pi$ (45) DWs. In the latter case, a $\pi$ DW that separates two antiparallel orientations of $\mathbf{P}$ in a unidirectionally rubbed cell splits into two surface disclinations. These disclinations are shifted with respect to each other in the plane of the sample, which results in a twist around the normal to the sample. The splitting of DWs into surface disclinations can also be observed in conventional nematics in unidirectionally rubbed cells [40]. Whether the defect represents a wall or two surface disclinations depends on the balance of elasticity and in-plane anchoring, as explained by Kléman (46). To verify whether the DWs in our experiments can be split into pairs of surface disclinations, we created in-plane shears by shifting one plate with respect to the other. The shear does not result in separation of the DW images, neither of the P-walls, nor the T-walls, which suggests that these walls are not split into surface disclinations. The issue of walls vs. disclinations is briefly discussed below to clarify the reason for conflicting reports in the literature (19, 45).

Consider a balance of surface anchoring and elasticity of an $N_F$ DW with a polarization realignment by $\pi$ from an in-plane easy direction, say, along the $x$-axis, to the direction (-$x$). In the $N_F$, in-plane anchoring is polar (12) and can be described by a potential $W(\varphi) = \frac{W_Q}{2}\sin^2\varphi - W_P(\cos\varphi - 1)$, where $\varphi$ is the angle between $\mathbf{P}$ and the $x$-axis, $W_P$ and $W_Q$ are polar and quadrupolar anchoring coefficients, respectively (19). Within the wall of a thickness $w$ along the $y$-axis, $\mathbf{P}$ realigns from $\varphi=0$ to $\varphi = \pi$ in the $(xy)$ plane perpendicular to the wall; the $z$-component of $\mathbf{P}$ is zero. The anchoring energy is $\left(W_P + \frac{W_Q}{4}\right)w$ per unit length of the wall. The elastic energy is $(\overline{K}/w^2)wh = \overline{K}(h/w)$, where $\overline{K}$ is some average of the bend and splay elastic constants. The equilibrium width of the domain wall is then $w = \sqrt{\frac{\overline{K}h}{W_P+\frac{W_Q}{4}}}$, while the wall energy is $F_w = 2\sqrt{\overline{K}h\left(W_P + \frac{W_Q}{4}\right)}$. This energy should be compared to the elastic energy estimate of two surface disclinations, which is roughly $F_d \approx 2\overline{K}$. A possible model of surface disclinations involves twists of opposite handedness along the normal to the cell and splay-bend in the bulk which yields a non-singular $\mathbf{P}$ orthogonal to the rubbing direction in the middle of the cell. Such a twist might expand to the rest of the sample, which would reduce the electrostatic energy, as recent experiments demonstrate (44). If the twist elastic energy and the electrostatic energy balance each other, the wall would split into two surface disclinations when the cell is thick, $h > \overline{K}/\left(W_P + \frac{W_Q}{4}\right)$. It is reasonable to expect that $\overline{K}$ is at least $10^{-11}$ N (37); according to Basnet et al. (19), in unidirectionally rubbed $N_F$ cells, $W_P \sim \frac{W_Q}{4} \sim 10^{-6}$ J/m$^2$. Therefore, the splitting of uncharged $\pi$ DWs into two disclinations is possible when $h > 5$ μm. Thinner cells exhibit DWs that do not split into disclinations, as described by Basnet et al. (19), while thicker (or strongly rubbed cells with higher anchoring strength) cells could feature DWs split into surface disclinations, as described by Yi et al. (45). In our experiments with the degenerate anchoring, $W_P = W_Q = 0$, and the analysis above suggests that the DW should be of infinite width, which contradicts the experimental observations in Figs. 1, 2, 4, and 5. The reason for the finite width of the DWs in the absence of in-plane anchoring is electrostatics: a nonzero $-\partial P_k/\partial k$ produces two oppositely charged sheets at the DW, and their mutual attraction stabilizes the finite $w$ against spreading favored by orientational elasticity (21, 26, 27), Fig.3(b).

### 2.4. Walls in freely suspended films.

The texture of freely suspended films of DIO show that the polarization **P** is parallel to the edges of the square opening, forming bend DWs along the diagonals, Fig.6. This arrangement is supported by the geometrical anchoring effect of the meniscus and by a strong tendency of **P** to align tangentially to any N$_F$ interface, which avoids a strong surface charge. Even a small tilt $\psi \sim 5°$ of **P** from the $xy$ plane of a bounding plate or an interface would produce a surface charge density $P_z \sim P_\psi \sim 6 \times 10^{-3}$ C m$^{-2}$, which is larger than the typical surface charge $(10^{-4} - 10^{-5})$ C m$^{-2}$ of adsorbed ions reported for the N (47, 48). Recent experiments (49) with curved capillaries filled with the N$_F$ and subject to a longitudinal electric field provide firm evidence of a strong tangential anchoring at the N$_F$ interfaces. As in the case of the parabolic DWs, the diagonal DWs are of a finite width, stabilized by the electrostatic effect. As described by the models below, disparity of the elastic constants is capable of reproducing the finite DWs width.

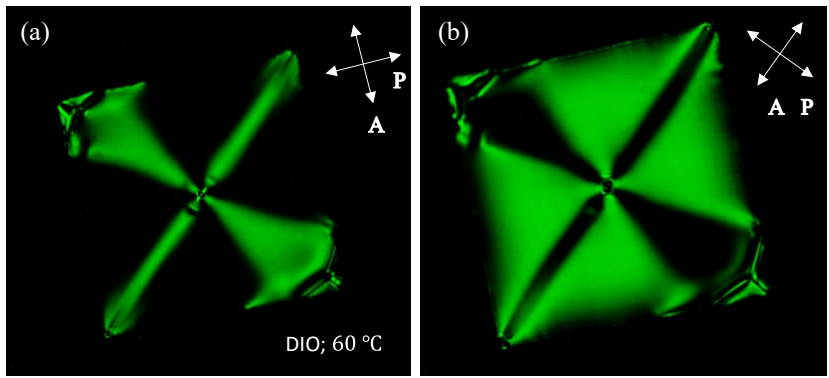

**Fig.6. Textures of freely suspended films of DIO.** Observations with (a) polarizers along the sides of the square hole; (b) polarizers oriented along the diagonals of the opening. Side length of the opening is 90 μm.

## 3. Modeling

In this section we will propose a mathematical model which sits within a standard approach to describe equilibrium configurations of nematic liquid crystals. A key aspect of our model is an assumption that the energetic cost of splay deformations far exceeds those of bend and twist. With such a theory in hand, we can capture a wide variety of experimentally observed morphologies. Success in this endeavor should also enable a reverse process of predicting material parameters of ferroelectric nematics on the basis of experimental observations.

As was already alluded to in the introduction, in modeling the experimental set-up, we propose an energy based on a vector-valued order parameter $\mathbf{P}$ representing the local polarization of the ferroelectric nematic sample. We then pursue analysis of an Oseen-Frank type of energy. However, in light of the fact that polarization may vanish in the neighborhood of defects or walls, we invoke a Landau-de Gennes (or Ginzburg-Landau) type of potential favoring unit vectors as opposed to a hard constraint that $|\mathbf{P}| = P_0$ everywhere in the sample. Here $P_0$ denotes a preferred value for the magnitude of the polarization vector.

It would also be natural to model the experiments discussed in this article using an energy based on both polarization and a nematic director, but as we shall indicate through numerous computational experiments in the subsequent section, our model based solely on $\mathbf{P} = (P^{(1)}, P^{(2)}, P^{(3)})$ already successfully captures an array of morphologies that emerge in the laboratory.

Given that the experiments are carried out on a thin domain, we take as our sample the set

$$\Omega = \{(x_1, x_2, x_3): 0 < x_1 < L,\ 0 < x_2 < L,\ -\frac{h}{2} < x_3 < \frac{h}{2}\},$$

where $L = O(1)$ and $0 < h \ll 1$. Then for $\mathbf{P}: \Omega \to \mathbb{R}^3$ we introduce the (dimensional) elastic energy

$$E(\mathbf{P}) := \int_\Omega \frac{K_1}{2}(\operatorname{div} \mathbf{P})^2 + \frac{K_2}{2}(\mathbf{P} \cdot \operatorname{curl} \mathbf{P})^2 + \frac{K_3}{2}|\mathbf{P} \times \operatorname{curl} \mathbf{P}|^2 \qquad (2)$$

$$+ \frac{\alpha}{4}(|\mathbf{P}|^2 - P_0^2)^2\, dx_1\, dx_2\, dx_3 + \frac{\gamma}{2} \int_{(x_1, x_2) \in \Omega, x_3 = \pm\frac{h}{2}} (P^{(3)})^4\, dx_1\, dx_2.$$

This is the analog of the Oseen-Frank energy for nematic liquid crystals written for polarization where we ignore electrostatic interactions. It is supplemented with the penultimate potential term fixing the preferred value for the magnitude of the polarization vector.

To explain the final surface anchoring term appearing in Eq. (2), we note that at the interface of an apolar nematic and an isotropic fluid, the only angular dependence that could enter the surface energy density is via a term proportional to $(\hat{\mathbf{n}} \cdot \hat{\mathbf{v}})^2$, where $\hat{\mathbf{n}}$ is the director and $\hat{\mathbf{v}}$ is the normal to the surface. For the anisotropic surface tension $\sigma_{NI}$ to yield a minimum at some "easy cone" $0 \leqslant \varphi_{eq} \leqslant \pi/2$, the term $(\hat{\mathbf{n}} \cdot \hat{\mathbf{v}})^2$ should be supplemented by higher-order terms, e.g. $\sigma_{NI} = \bar{\gamma}(\hat{\mathbf{n}} \cdot \hat{\mathbf{v}})^2 + \frac{\gamma}{2}(\hat{\mathbf{n}} \cdot \hat{\mathbf{v}})^4 + const$, which can be rewritten as

$$\sigma_{NI} = \sigma_0 + \frac{\gamma}{2}\left[(\hat{\mathbf{n}} \cdot \hat{\mathbf{v}})^2 - (\hat{\mathbf{n}}_{eq} \cdot \hat{\mathbf{v}})^2\right]^2,$$

as long as $-1 \leq \bar{\gamma}/\gamma \leq 0$. Here $\hat{\mathbf{n}}_{eq}$ is the "easy axis" making the equilibrium angle $\varphi_{eq}$ with $\hat{\mathbf{v}}$, defined from $\cos^2 \varphi_{eq} = -\frac{\bar{\gamma}}{\gamma}$. If $\hat{\mathbf{n}}_{eq}$ is tangential to the interface, then $\varphi_{eq} = \pi/2$ and $\hat{\mathbf{n}}_{eq} \cdot \hat{\mathbf{v}} = 0$, thus

$$\sigma_{NI} = \sigma_0 + \frac{\gamma}{2}(\hat{\mathbf{n}} \cdot \hat{\mathbf{v}})^4.$$

A similar consideration is valid for the ferroelectric nematic with $\mathbf{P}$ replacing $\hat{\mathbf{n}}$ and $\hat{\mathbf{v}} = (0,0,1)$ as in Fig.7. This justifies the term with $\frac{\gamma}{2}(P^{(3)})^4$ describing the zenithal surface anchoring.

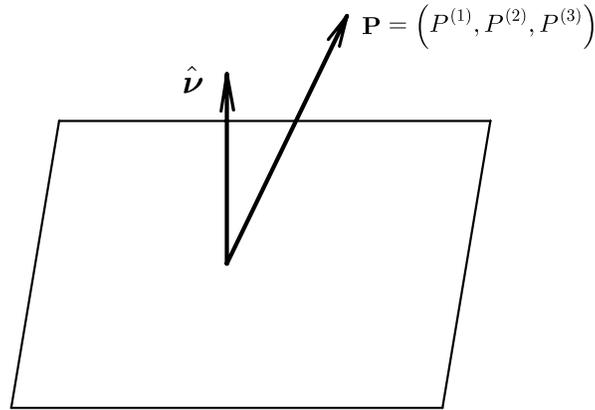

**Fig.7. Boundary schematics.** Here $\mathbf{P}$ is the polarization vector and $\hat{\mathbf{v}}$ is the surface normal.

By setting

$$K_2 = K_3 = K_1 - M = K, \quad K_2 + K_4 = 0, \quad where \ M > 0,$$

we arrive at the following simplified expression for the elastic energy

$$E(\mathbf{P}) := \int_\Omega \frac{M}{2}(\operatorname{div}\mathbf{P})^2 + \frac{K}{2}|\nabla\mathbf{P}|^2 + \frac{\alpha}{4}(|\mathbf{P}|^2 - P_0^2)^2 \, dx_1\, dx_2\, dx_3$$
$$+ \frac{\gamma}{2}\int_{(x_1,x_2)\in\Omega, x_3=\pm\frac{h}{2}} (P^{(3)})^4 \, dx_1\, dx_2.$$

This version of the model captures the experimentally observed features of ferroelectric nematics. Since a key feature of the ferroelectric nematics under consideration here is the relatively high energetic cost of splay, we assume that $M \gg K$ where we note that in the second term above we are effectively folding the cost of twist and bend into one equal constant term. Because of the degenerate planar anchoring, the angle between the polarization and any bounding surface is essentially zero and, therefore, the anchoring strength $\gamma \gg 1$.

Taking a characteristic diameter of $\Omega$ to be, say $L$, we identify two small dimensionless parameters:

$$\delta := \frac{h}{L} \quad \text{and} \quad \varepsilon := \frac{\sqrt{K/\alpha}}{LP_0}, \tag{3}$$

with two more dimensionless parameters

$$\mu := \frac{M}{LP_0\sqrt{K\alpha}} \quad \text{and} \quad \Gamma := \frac{\gamma P_0}{\sqrt{K\alpha}}, \tag{4}$$

whose size will be selected in the sequel. Here the smallness of $\delta$ and $\varepsilon$ follow respectively from the high aspect ratio of the film and smallness of the ferroelectric nematic correlation length, $\xi_n = \frac{\sqrt{K/\alpha}}{P_0}$, relative to the width $L$ of the film.

Then rescaling the spatial variables via

$$x := \frac{x_1}{L}, \quad y := \frac{x_2}{L} \quad \text{and} \quad z := \frac{x_3}{h},$$

and polarization by

$$\widetilde{\mathbf{P}} := \frac{1}{P_0}\mathbf{P}$$

and finally dividing $E$ by $hL\sqrt{K\alpha}$ and dropping tildes, we arrive at a dimensionless expression for the energy (which we still denote by $E$):

$$E(\mathbf{P}) = \int_{\Omega_0 \times (-\frac{1}{2},\frac{1}{2})} \frac{\mu}{2}\left(P_x^{(1)} + P_y^{(2)} + \frac{1}{\delta}P_z^{(3)}\right)^2 + \frac{\varepsilon}{2}\left(|\mathbf{P}_x|^2 + |\mathbf{P}_y|^2 + \frac{1}{\delta^2}|\mathbf{P}_z|^2\right)$$
$$+ \frac{1}{4\varepsilon}(|\mathbf{P}|^2 - 1)^2 \, dx\, dy\, dz + \frac{\Gamma}{\delta}\int_{\Omega_0 \times \{\pm\frac{1}{2}\}} (P^{(3)})^4 \, dx\, dy, \tag{5}$$

for a polarization vector $\mathbf{P}$ defined for $(x,y) \in \Omega_0$ and $z \in (-\frac{1}{2},\frac{1}{2})$. Here $\Omega_0$ is a square of unit side length and the subscripts $x, y$, and $z$ denote derivatives with respect to these variables.

We seek minimizers of $E$ with bounded energy. Because $\delta$ and $\varepsilon$ are small, while $\mu$ and $\Gamma$ are both $O(1)$, we observe that dependence of $\mathbf{P}$ on the variable $z$, the nonvanishing third component of $\mathbf{P}$ and deviations of $|\mathbf{P}|$ from $1$ all incur very high energy cost. In light of the first fact, in what follows we make a further simplification that the polarization $\mathbf{P}$ is independent of $z$, leading to the reduced energy

$$E_r(\mathbf{P}) = \int_{\Omega_0 \times (-\frac{1}{2},\frac{1}{2})} \frac{\mu}{2}\left(P_x^{(1)} + P_y^{(2)}\right)^2 + \frac{\varepsilon}{2}\left(|\mathbf{P}_x|^2 + |\mathbf{P}_y|^2\right) \\ + \frac{1}{4\varepsilon}(|\mathbf{P}|^2 - 1)^2 \, dx \, dy + \frac{2\Gamma}{\delta}\int_{\Omega_0}(P^{(3)})^4 \, dx \, dy, \qquad (6)$$

Now, the behavior of a configuration that minimizes Eq. (5) is dictated by relative sizes of $\varepsilon$ and $\delta$.

Suppose first that $\delta \ll \varepsilon$. In this case we can conclude that the last term in Eq. (5) dominates unless the third component of $\mathbf{P}$ vanishes. Therefore, we can impose the condition that $\mathbf{P}$ lies in $xy$-plane. Under these assumptions, to leading order we find that the resulting energy $\mathcal{E}_\varepsilon$ can be described through a two-component vector field $\mathbf{p} = (p^{(1)}(x,y), p^{(2)}(x,y))$ via

$$\mathcal{E}_\varepsilon(\mathbf{p}) \sim \int_{\Omega_0} \frac{\mu}{2}(\text{div } \mathbf{p})^2 + \frac{\varepsilon}{2}|\nabla \mathbf{p}|^2 + \frac{1}{4\varepsilon}(|\mathbf{p}|^2 - 1)^2 \, dx \, dy. \qquad (7)$$

The assumption of $z$-independence aligns with the experimental observations carried out in thin samples sandwiched between two interfaces. However, interestingly enough, numerical simulations and formal asymptotics indicate that this reduction from Eq. (6) to Eq. (7) is far from being mathematically straightforward and thus will not be discussed here.

Now, instead let $\varepsilon \ll \delta$. Then the integral of $\frac{1}{4\varepsilon}(|\mathbf{P}|^2 - 1)^2$ will be the largest contributor to the energy $E_r$ so that we are justified to assume that $|\widehat{\mathbf{P}}| \equiv 1$ in $\Omega_0$. Writing $P^{(3)}$ in terms of other components of $\mathbf{P}$ then gives

$$\left(P^{(3)}\right)^4 = \left(1 - \left(P^{(1)}\right)^2 - \left(P^{(2)}\right)^2\right)^2 = (1 - |\mathbf{p}|^2)^2.$$

We also have

$$P_x^{(1)} + P_y^{(2)} = \text{div } \mathbf{p},$$

because $\mathbf{p}$ is independent of $z$. Substituting these expressions into (6), we obtain the energy

$$\mathcal{F}_\varepsilon(\mathbf{p}) \sim \int_{\Omega_0} \frac{\mu}{2}(\text{div } \mathbf{p})^2 + \frac{\varepsilon}{2}\left(|\nabla \mathbf{p}|^2 + \left|\nabla P^{(3)}\right|^2\right) + \frac{2\Gamma}{\delta}(|\mathbf{p}|^2 - 1)^2 \, dx \, dy, \qquad (8)$$

where $|\mathbf{p}|^2 + \left(P^{(3)}\right)^2 \equiv 1$ in $\Omega_0$.

We observe that the expressions for Eq. (7) and Eq. (8) are mathematically very similar even though the potential terms originate from two unrelated sources. The differences arise from the presence of the gradient of the third component in Eq. (8) and different relationships between the parameters. From the physical perspective, the energy in Eq. (8) allows for twist deformation—in particular, within a wall—while Eq. (7) only permits splay and bend. In the next subsection we discuss asymptotics for Eq. (7). Then, in the following section devoted to numerics, we will present examples of energy-minimizing configurations for both Eq. (7) and Eq. (8).

### 3.1. Asymptotic analysis

In this section we indicate how to mathematically analyze energy minimizing configurations for Eq. (7), taking advantage of the smallness of $\varepsilon$. In particular, we describe analytical techniques for constructing elastic walls for given anchoring conditions on the two-dimensional polarization $\mathbf{p}$. This subsection aims to provide the mathematical justification for the numerical results presented in the next section.

The energy in Eq. (7) penalizing splay over bend has been analyzed in (50). The most salient observation in (50) is that, as $\varepsilon \to 0$, the minimization problem (7) approaches a sum of bulk splay cost and a wall cost given by

$$\mathcal{E}_0(\mathbf{p}) := \int_{\Omega_0} \frac{\mu}{2}(\text{div } \mathbf{p})^2 \, dx \, dy + \frac{1}{3}\int_{J_\mathbf{p}} |\mathbf{p}_+ - \mathbf{p}_-|^3 \, ds, \qquad (9)$$

where $\mathbf{p}$ has prescribed values on $\partial\Omega_0$ and satisfies $|\mathbf{p}| = 1$ everywhere in $\Omega_0$. The symbol $J_\mathbf{p}$ represents the domain wall, that is, a curve across which $\mathbf{p}$ jumps in order to save on the cost of splay, and $\mathbf{p}_+$ and $\mathbf{p}_-$ denote the values of the polarization on either side of the wall.

A mathematical subtlety we wish to highlight with regard to the energy $\mathcal{E}_0$ is that eligible vector fields $\mathbf{p}$ that exhibit jump discontinuities across such a curve $J_\mathbf{p}$ must nonetheless respect the integration by parts formula (i.e. the Divergence Theorem). This induces a requirement that the normal component of **p** remains continuous across the wall. Since the potential term in Eq. (7) forces the polarization field **p** to be of a unit magnitude away from the wall for $\varepsilon \ll 1$, it follows that the tangential component of **p** simply switches sign on either side of the wall. Thus, we see that the physical requirement that a domain wall be uncharged, as discussed in the introduction, is manifested in our model through an application of integration by parts.

A simple example that illustrates the utility of this continuity condition is in order. Suppose that a wall separates two distinct states with zero splay, for example, a state where $\mathbf{p} \equiv const$ and a state where **p** has a circular vortical pattern. In view of Eq. (9), then all of the energy of such a configuration will be concentrated on the wall. Then the placement of the wall is dictated by the requirement that across it, the tangential component of polarization switches sign.

To be specific, suppose that the constant state is $\mathbf{p} = (1,0)$ and the divergence-free circular vortical state is

$$\mathbf{p} = (-\sin\theta, \cos\theta). \tag{10}$$

Let us describe the wall in terms of its distance, say $\rho(\theta)$, from the origin which serves as the center of the vortex. That is, as a function of the polar angle $\theta$, suppose the wall is given by

$$\theta \mapsto \rho(\theta)(\cos\theta, \sin\theta).$$

Then a tangent vector to the wall is given by

$$\mathbf{T}(\theta) := \rho'(\theta)(\cos\theta, \sin\theta) + \rho(\theta)(-\sin\theta, \cos\theta),$$

and the condition of sign-switching tangential component and continuous normal component can be expressed as

$$(1,0) \cdot \mathbf{T} = -(-\sin\theta, \cos\theta) \cdot \mathbf{T}.$$

This equation simplifies to the separable ODE

$$\frac{\rho'}{\rho} = \frac{\sin\theta - 1}{\cos\theta},$$

which can be readily solved to find that the wall is given by

$$\rho(\theta) = \frac{C}{1+\sin\theta} \quad for\ some\ constant\ C. \tag{11}$$

Observing that Eq. (11) is the polar equation of a parabola, we see that a these two divergence-free states must be separated by a parabolic domain wall, also known as a $P-$wall, as already discussed in the introduction and in the experimental part. Numerical simulations conducted through minimization of $\mathcal{E}_\varepsilon$ for $\varepsilon$ small confirm that this geometry emerges for appropriate boundary conditions. What is more, this arrangement conforms with the expeccted emergence of conics dicussed in the introduction and is consistent with experimental observations as well, giving support for the validity of the model (cf. (20), Fig. 7).

Another configuration consistent with splay-free bulk, and one that emerges experimentally as well, is the appearance of domain walls in a triple junction configuration. In the simplest scenario, consider three rays, meeting at the origin and separated from each other by an angle of $120°$, i.e., at angles $0°$, $120°$ and $240°$ with the $x$-axis. Within each $120°$ sector, place a uniform state so that the director makes an angle of $150°$ with the $x$-axis in the first sector, an angle of $-90°$ with the $x$-axis in the second sector and an angle of $30°$ with the $x$-axis in the third sector. Just as in the previous example, the entire energy of this state is concentrated on the walls and the configuration respects the continuity condition of the normal component of polarization across all walls. Another example comes from exchanging the constant states from the previous example with three circular vortices. Once again, all of these configurations may be observed experimentally.

One can also use Eq. (9) to construct walls in more complicated settings where, for example, the divergence does not vanish on either side of the walls; again see (50). For this pursuit, we observe that a minimizer of $\mathcal{E}_0$ satisfies the criticality condition

$$\mathbf{p}^\perp \cdot \nabla (\operatorname{div} \mathbf{p}) = 0 \quad in \ \Omega_0 \setminus J_\mathbf{p}, \quad \text{where} \ \mathbf{p}^\perp = (-p^{(2)}, p^{(1)}), \tag{12}$$

along with the requirement $|\mathbf{p}| = 1$. Writing $\mathbf{p}$ locally as $\mathbf{p}(x,y) = (\cos\theta(x,y), \sin\theta(x,y))$ and defining the scalar

$$\eta := \operatorname{div} u, \tag{13}$$

one has that Eqs. (11)-(12) is equivalent to the following system of partial differential equations for the two scalars $\theta$ and $\eta$:

$$-\sin\theta \ \theta_x + \cos\theta \ \theta_y = \eta, \tag{14}$$

$$-\sin\theta \ \eta_x + \cos\theta \ \eta_y = 0. \tag{15}$$

This system of first order partial differential equations has a fairly simple solution obtainable by the method of characteristics, cf.(51). Starting from any initial curve in $\Omega_0$ parametrized via $s \mapsto (x_0(s), y_0(s))$ along which $\theta$ and $\eta$ take known values $\theta_0(s)$ and $\eta_0(s)$ respectively, the characteristic curves parametrized by $t$ are given by

$$x(s,t) = \frac{1}{\eta_0(s)}[\cos(\eta_0(s)t + \theta_0(s)) - \cos\theta_0(s)] + x_0(s),$$

$$y(s,t) = \frac{1}{\eta_0(s)}[\sin(\eta_0(s)t + \theta_0(s)) - \sin\theta_0(s)] + y_0(s),$$

for each $s$ whenever $\eta_0(s) \neq 0$. The corresponding solutions $\theta$ and $\eta$ are then given by

$$\theta(s,t) = \eta_0(s)t + \theta_0(s), \quad \eta(s,t) = \eta_0(s),$$

so that the characteristics are circular arcs of curvature $\eta_0(s)$ and carry constant values of the divergence. In case the divergence vanishes somewhere along the initial curve, i.e. $\eta_0(s) = 0$, then the characteristic is a straight line.

Examples of employing this method can be found in (50). In particular, consider a rectangular domain where the polarization points to the right on the top, to the left on the bottom and is periodic on the vertical sides of the rectangle. For certain parameter regimes, minimizing configurations for the energy in Eq. (7) can be shown to be characterized by the presence of walls, as shown in Fig. 8. The respective level plots of the angle of inclination of polarization with the $x$-axis, and of the divergence of polarization are shown in Fig. 9. Note that the divergence concentrates along two nearly parallel and close stripes that embrace the DW, Fig.9a; we associate these stripes with a nonzero bound charge $-\partial P_k/\partial k$ which is maximized near the wall and changes sign from positive to negative as one crosses the wall.

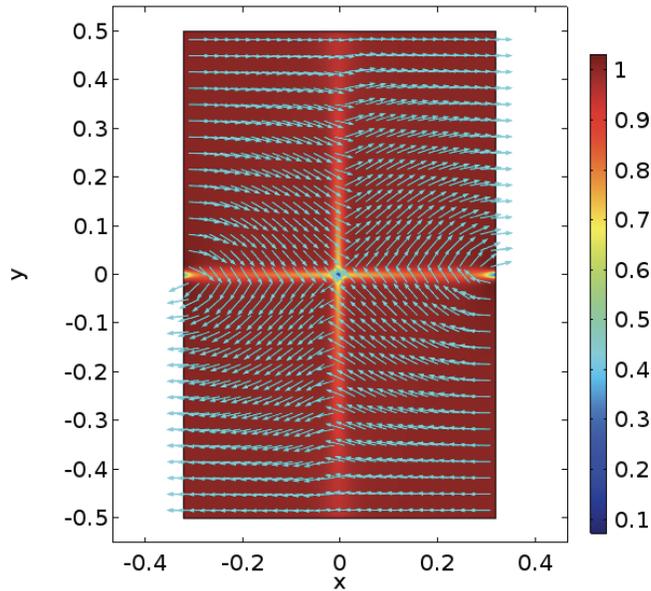

**Fig. 8:** A solution **p** of the Euler-Lagrange equation associated with the energy functional in Eq. (7) in the rectangle $(-0.3, 0.3) \times (-0.5, 0.5)$ subject to periodic boundary conditions on $\{-0.3, 0.3\} \times (-0.5, 0.5)$ and assuming that $\mathbf{p}(x, \pm 0.5) = (\pm 1, 0)$. Here $\mu = 1$ and $\varepsilon = 0.005$. Both **p** and $|\mathbf{p}|$ are shown.

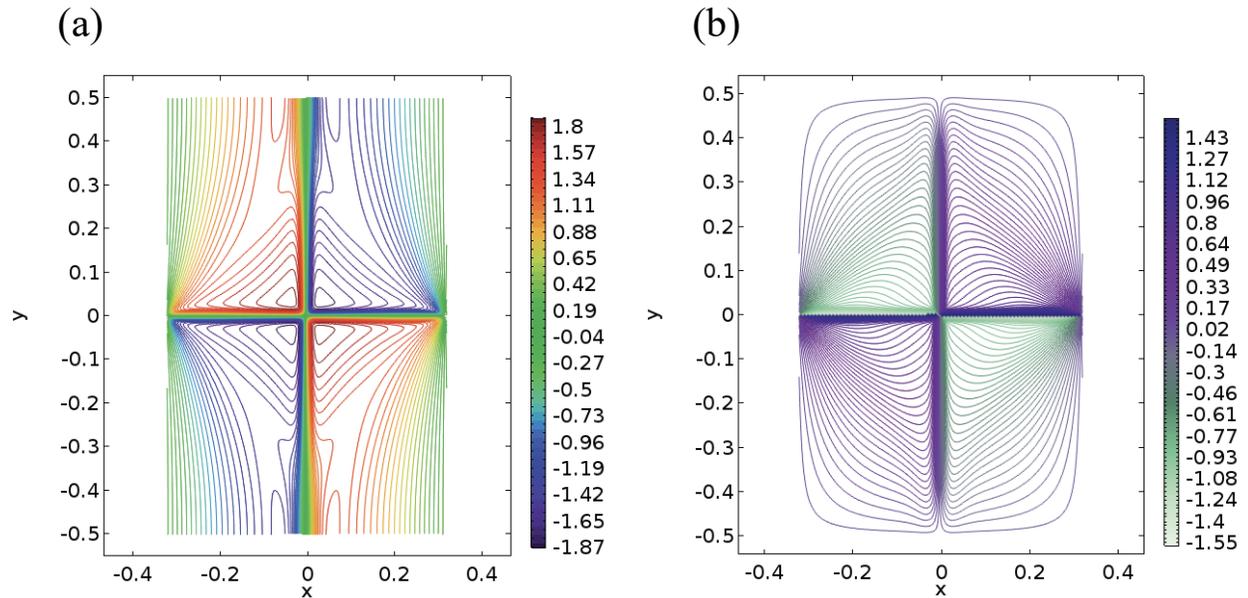

**Fig. 9:** Level curves for the divergence of **p** (a) and the angle $\theta$ (b), where $\mathbf{p} = (\cos\theta, \sin\theta)$ is depicted in Fig. 7.

This configuration was obtained numerically in (50) using the finite elements package COMSOL (52) by solving the full system of partial differential equations that describe the critical points of Eq. (7). In this case, the walls have a simple shape of a cross, but the morphology of the polarization field is fairly complex. Nonetheless, the same solution can be constructed analytically by using the method of characteristics described above. The corresponding plots are shown in Fig. 10 for the top right quarter of the rectangle and they clearly have the same behavior as the numerically derived result in Fig. 9.

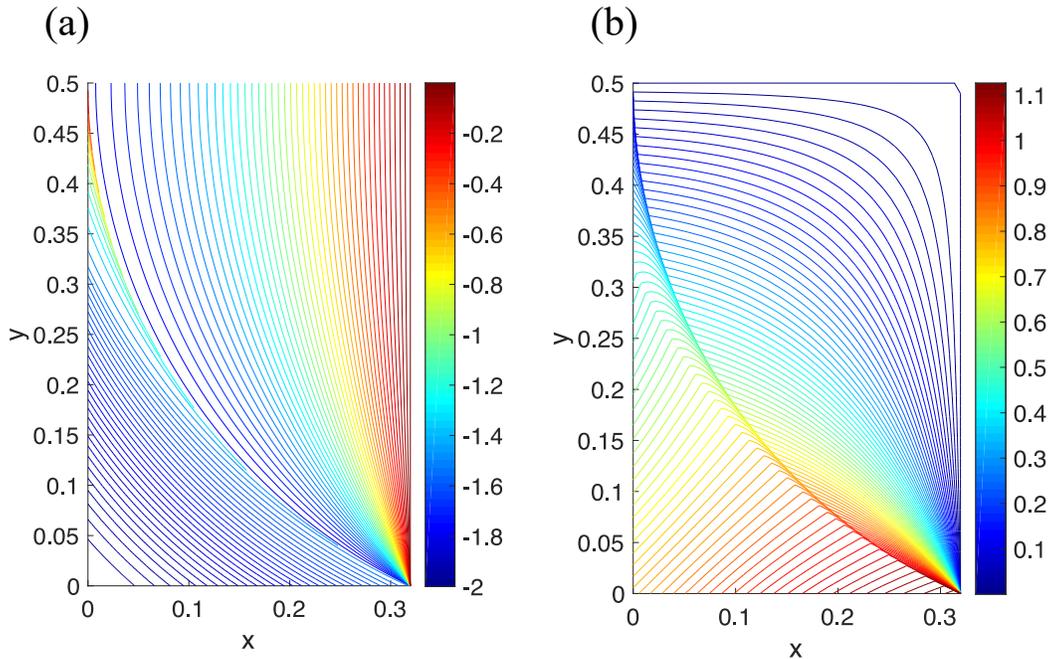

**Fig. 10**: Level curves for the divergence of $\mathbf{p}$ (a) and the angle $\theta$ (b), where $\mathbf{p} = (\cos\theta, \sin\theta)$ is a solution obtained using characteristics.

4. **Numerics**

In this section, we demonstrate that the minimizers of the two-dimensional energy in Eq. (7) correctly describe behavior of polarization in real ferroelectric nematic films. The observed domain wall structures are triggered by the boundary conditions, presence of impurities and inclusions.

To benchmark our model, we will begin by looking for an optimal configuration of Eq. (7) in a rectangular domain with boundary conditions corresponding to a circular polarization on the bottom edge of the rectangle and the constant state $\mathbf{p} = (1,0)$ on the other three sides.

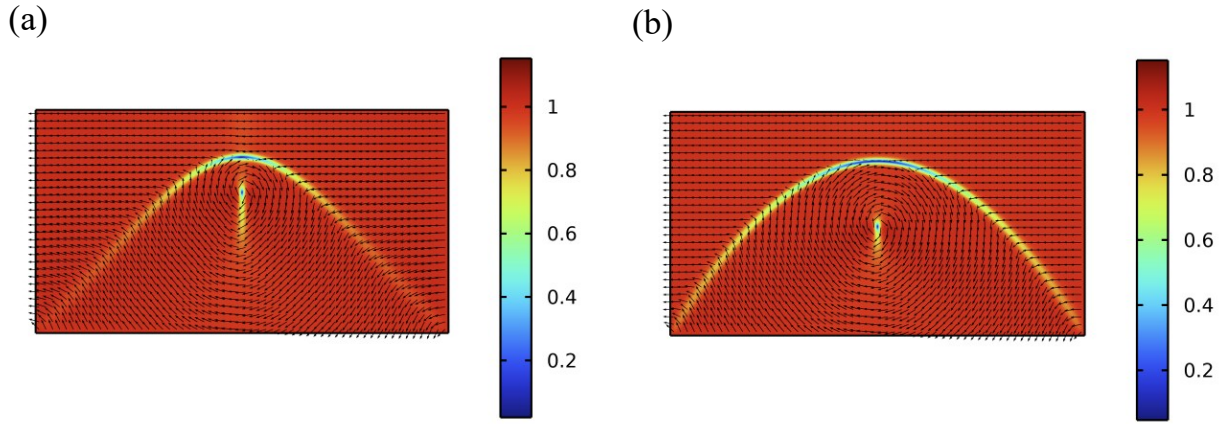

**Fig. 11**: A parabolic wall in a ferroelectric nematic film induced by the boundary conditions. The arrows represent the two-dimensional polarization vector $\mathbf{p}$ and the color corresponds to $|\mathbf{p}|$. Here $\varepsilon = 0.03$, $\mu = 10$ (a) and $\varepsilon = 0.03$, $\mu = 1000$ (b).

It is well known that the energy minimizer in this case is characterized by the presence of a parabolic wall, Fig. 11 that indeed forms in the domain during the gradient descent for Eq. (7). This outcome also corresponds to the analysis in the previous section, cf. Eq. (11).

Next, we will consider inclusions that have sizes comparable to the thickness of the film. To this end, consider the situation shown in Fig. 12.

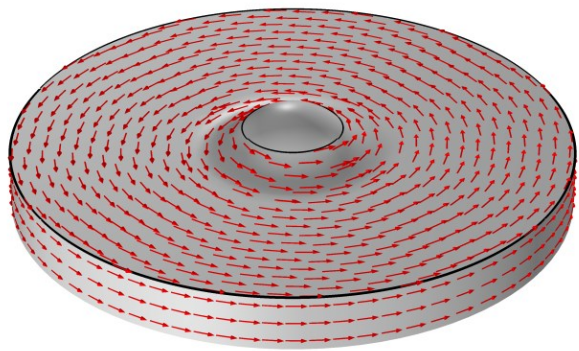

**Fig. 12.** A meniscus forming around an inclusion orients the polarization field $\mathbf{P}$.

Here, a spherical inclusion penetrates the top and the bottom of the film at the interfaces with air and glycerol and therefore a meniscus forms on each boundary. As alluded to previously, because the polarization vector $\mathbf{P}$ wants to remain parallel to these interfaces, in order to minimize the elastic energy in three dimensions $\mathbf{P}$ orients along the normal to the thickness gradient.

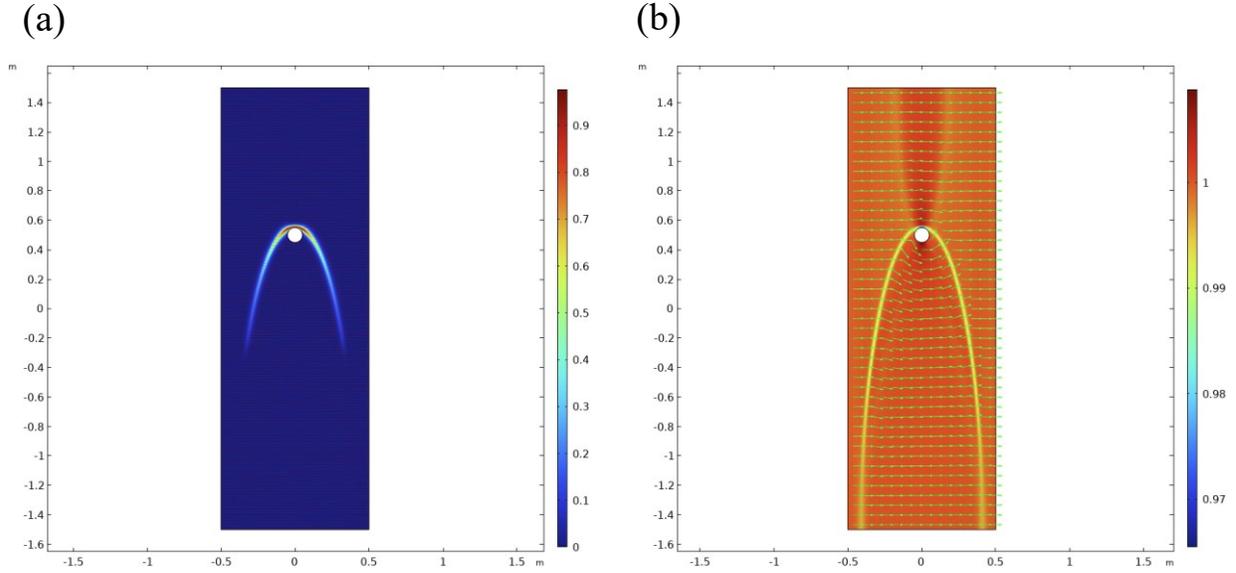

**Fig. 13.** The component $P^{(3)}$ of the polarization vector (a); the projection $\mathbf{p}$ of polarization $\mathbf{P}$ onto the plane of the film. The color corresponds to $|\mathbf{P}|$ (b). Here $\mu = 6$, $\varepsilon = 0.002$, $\Gamma = 1$, and $\delta = 0.12$.

Therefore, we expect that the polarization should be pinned in a circular pattern around the inclusion on both surfaces of the film, cf. Fig. 12. In order to model a polarization field in a ferroelectric nematic film with inclusions within the $2D$-framework, we can then excise a disk corresponding to the inclusion along with the meniscus and impose tangential anchoring on the boundary of the disk.

Now we consider a rectangular domain with a hole representing an excised disk around an inclusion. We impose constant boundary conditions $\mathbf{P} = (1,0,0)$ on the boundary of the rectangle and tangential boundary conditons, $\mathbf{P} = (-\sin\theta, \cos\theta, 0)$ on the boundary of the disk for a polar angle $\theta$ with respect to the center of the disk. We then find the energy-minimizing configuration of polarization by using steepest descent for the energy Eq. (8), as shown in Fig. 13.

We observe a parabolic downward-pointing wall with a faint upward parabolic "ghost" wall. These features have also been observed in experiments, Fig.4. Note that in this case the elastic deformation at the tip of the wall is dominated by twist around the axis in the $xy$-plane. As already discussed, such a twist under the condition of $z$-independent polarization might be prevented by electrostatics at most interfaces. Experiments suggest that the details of the fine structure of domain walls involve all three types of deformations, including variations along the $z$-axis, which reconcile the twist in the bulk with tangential anchoring at the boundaries (19). However, we do not exclude a possibility that in the future a homeotropic or strongly tilted anchoring might be achieved at some $N_F$-substrate interfaces, in which case such a twist across the entire thickness of an $N_F$ slab could be observed.

Figures 14 and 15 confirm that our modeling approach also produces the correct behavior in the system consisting of two particles imbedded in a ferroelectric nematic film. Figures 1(d-g) show the corresponding experimental image. The plot in the Figs. 14 and 15 are, respectively, minimizers of the energy (8) and (7) in the region exterior to these disks. It is apparent that, while the wall morphologies in the two figures match experimental observations, there are some subtle differences, e.g., the secondary walls are present only in Fig. 15.

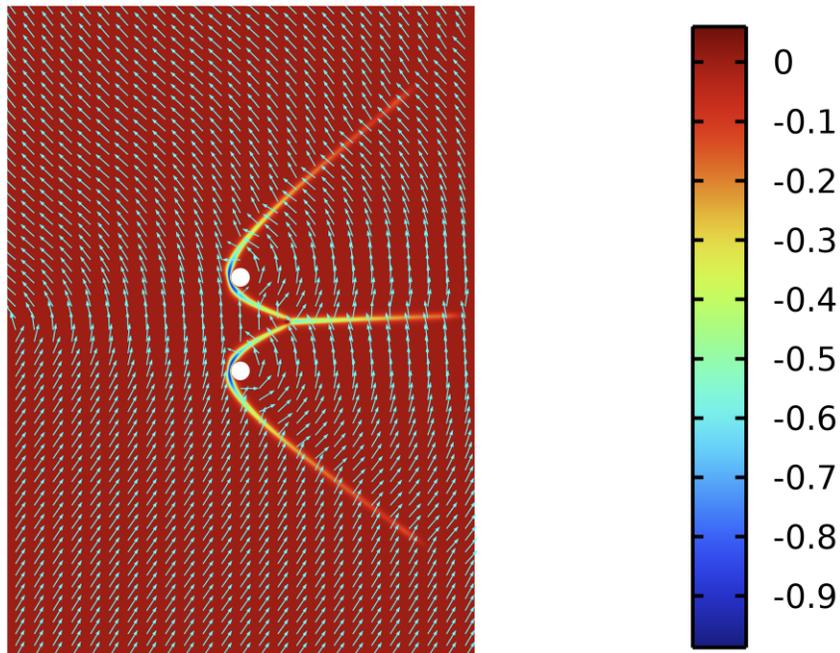

**Fig. 14:** Morphology of a ferroelectric nematic film with two inclusions: an energy minimizing configuration for the energy Eq. (8). Here $\mu = 8$, $\varepsilon = 0.004$, $\Gamma = 1$, and $\delta = 0.004$. The color

indicates the value of $P^{(3)}$, while the arrows represent the projection $\mathbf{p}$ of polarization onto the plane of the film.

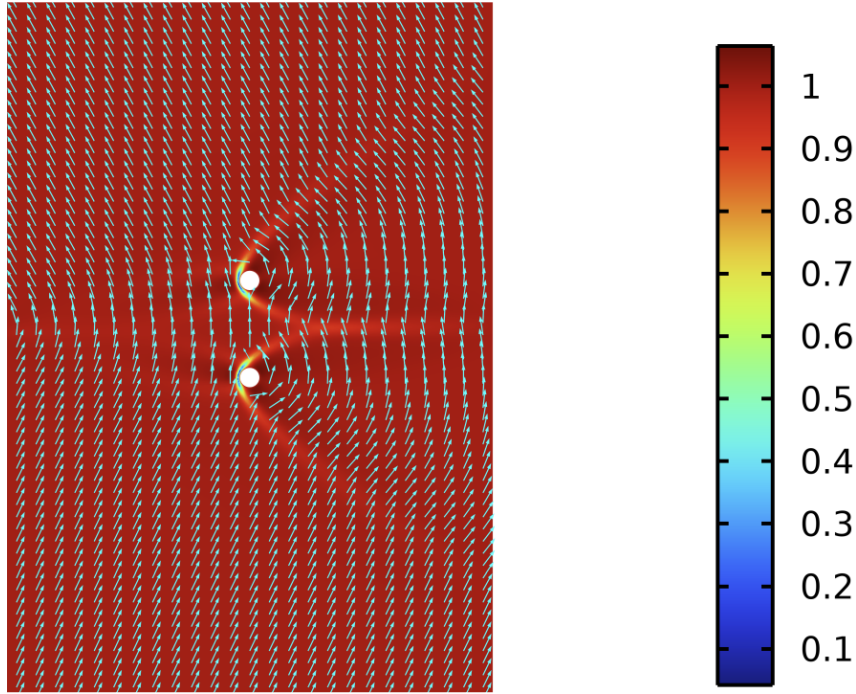

**Fig. 15.** Morphology of a ferroelectric nematic film with two inclusions: an energy minimizing configuration for the energy Eq. (7). Here $\mu = 8$ and $\varepsilon = 0.004$. The color indicates the value of $|\mathbf{P}|$, while the arrows represent the polarization $\mathbf{P} = \left(P^{(1)}, P^{(2)}, 0\right)$. Note that here $\mathbf{P} = \mathbf{p}$.

The model Eq. (7) is based on a two-component polarization vector and therefore the walls in Fig. 15 are of the bend-splay type. On the other hand, because the model Eq. (8) involves three components, the corresponding wall structure is allowed to exhibit twist and indeed, this is what occurs in Fig. 14 as can be discerned from the deviations of the third component of $\mathbf{P}$ away from zero in the interior of the wall. In Fig. 16 we show a similar configuration for four inclusions.

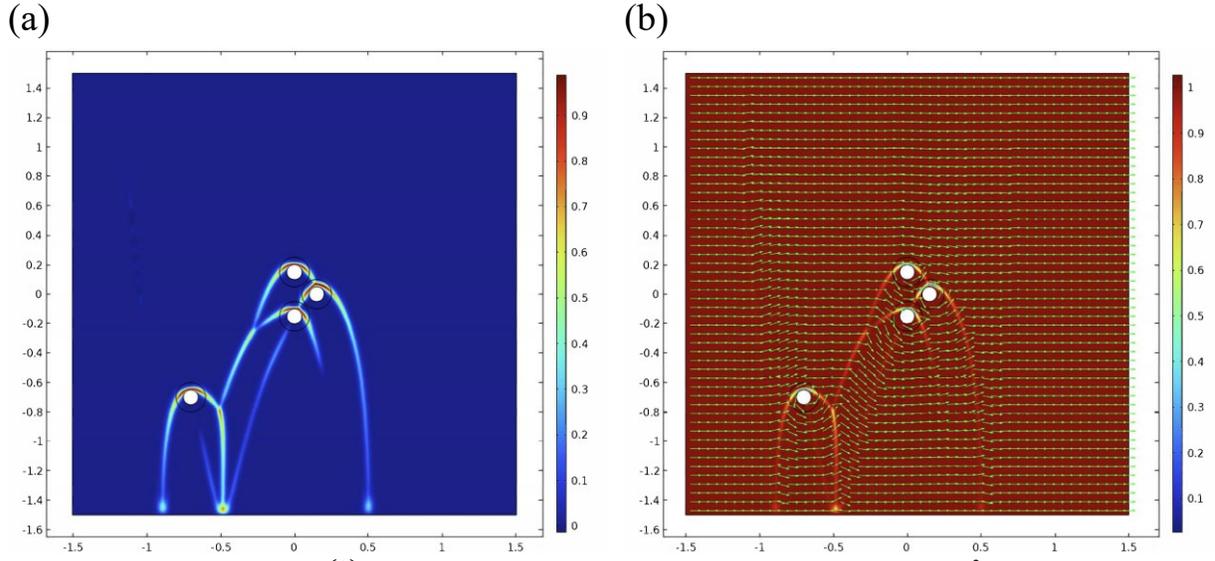

**Fig. 16**. The component $P^{(3)}$ of the polarization vector (a); the projection $\hat{\mathcal{P}}$ of polarization **P** onto the plane of the film. The color corresponds to $|\mathbf{P}|$ (b). Here $\mu = 6$, $\varepsilon = 0.002$, $\Gamma = 1$, and $\delta = 0.12$.

The plot on the right in Fig. 17 is obtained by using steepest descent of the energy Eq. (8) starting from initial data with regions of both clockwise and counterclockwise twist of the polarization. This plot shows a single inclusion with an associated parabolic wall. However there is a difference with Fig. 13 in that the left half of the wall has positive twist, while its right half has negative twist.

These halves are separated by a point defect at the tip of the parabola. Similar structures can be observed in experimental images in Fig. 1(d). The same phenomenon can be induced by considering a strip $(-1,1) \times (-1/2, 1/2)$ with antipodal boundary conditions on horizontal components of the strip, i.e., $\mathbf{P}(x, -1/2) = -\mathbf{P}(x, 1/2) = (1,0,0)$.

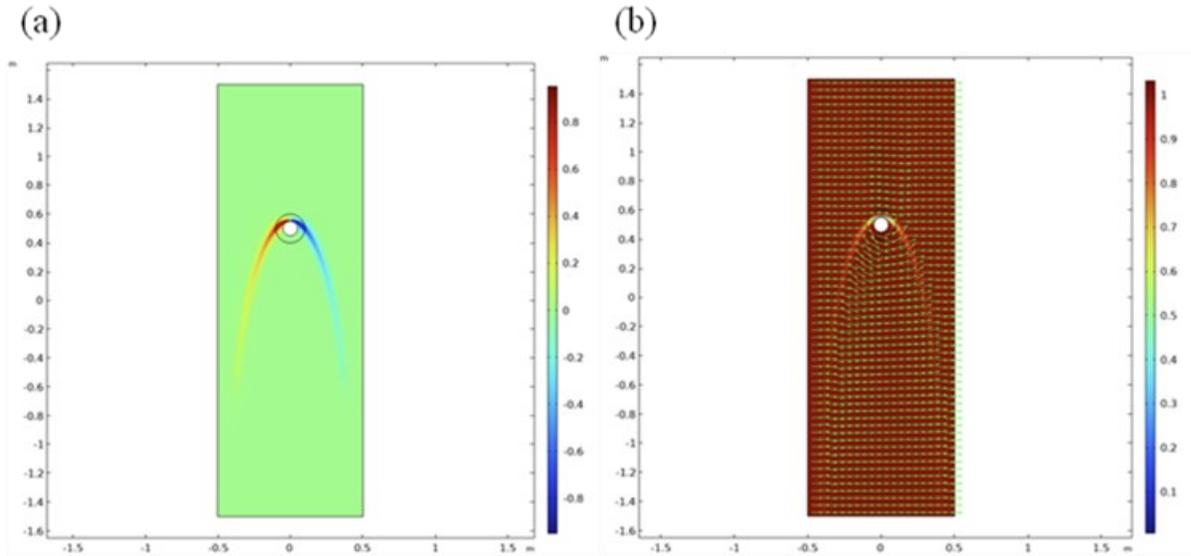

**Fig. 17.** The component $P^{(3)}$ of the polarization vector (a); the projection **p** of polarization **P** onto the plane of the film. The color corresponds to $|\mathbf{P}|$ (b). Here $\mu = 6$, $\varepsilon = 0.002$, $\Gamma = 1$, and $\delta = 0.12$.

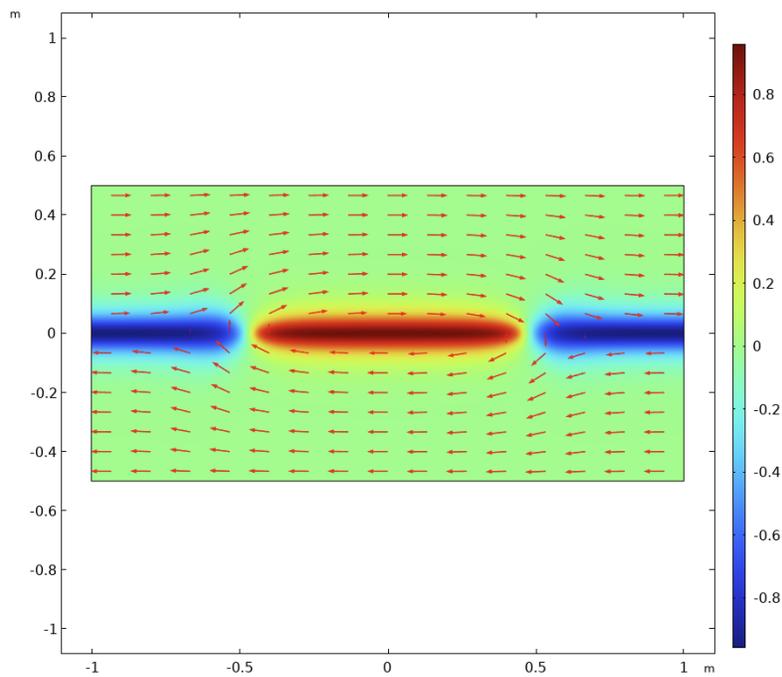

**Fig. 18**: An energy minimizing configuration in a periodic strip with antipodal orientation of the polarization vector on the top and the bottom. The color represents the value of $P^{(3)}$. Here $\mu = 0.2$, $\varepsilon = 0.008$, $\Gamma = 1$, and $\delta = 0.2$.

Suppose that the polarization field satisfies periodic boundary conditions on the vertical sides of the strip. Starting from initial data containing twist of both signs and using gradient descent leads to a local minimum of the energy in Eq. (8) with a wall along the axis $y=0$ such that the twist alternates as one moves along the wall, Fig. 18. Once again, the intervals of opposite twist are separated by point defects.

Finally, to model freely suspended films of DIO, we consider the energy (7) over a square domain where the direction of the polarization vector on the boundary is parallel to the boundary itself. The results are shown in Fig. 19 and demonstrate formation of bend-splay domain walls along the diagonals of the square, which one can compare favorably to the experimental findings in Fig. 6.

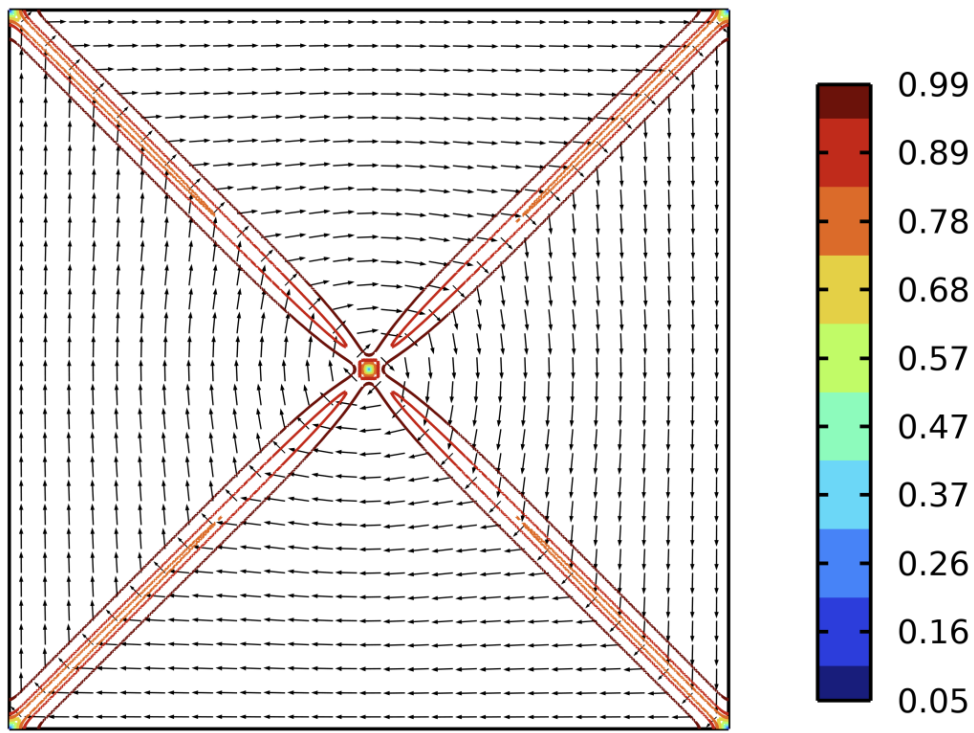

**Fig. 19:** An energy minimizing configuration in a square domain with the polarization vector parallel to the boundary. The arrows represent the polarization $\mathbf{p}$ and the color corresponds to the contours of $|\mathbf{p}|$. Here $\mu=10$ and $\varepsilon=0.03$.

## 6. Conclusions

As compared to their solid counterparts, ferroelectric nematics show a rich variety of polarization domain structures not constrained by crystallographic axes. The polarization patterns $\mathbf{P}(x,y,z)$ are controlled by the avoidance of bound charge associated with the divergence of the spontaneous polarization and by surface anchoring; the anchoring shows a strong tendency of $\mathbf{P}$ to be tangential to the $N_F$ interfaces. Thin $N_F$ films with degenerate in-plane anchoring exhibit circular domains of $\mathbf{P}$ separated from the domains of a uniform $\mathbf{P}$ by DWs in the shapes resembling parabolae, which help to reduce the bound charge. The DWs are of finate width, even when there is no in-plane (azimuthal) anchoring, which is the case of the films on the glycerol surface and in freely suspended films. The finite width of the DWs is stabilized by electrostatics.

Except near the tip of the DW in azimuthally-degenerate slabs, the eccentricity is close to 1, hence the name "P-wall". The P-wall structure is complex, with the bend and splay of $\mathbf{P}$ in the $(x,y)$ plane of the film and a twist of $\mathbf{P}$ along the normal $z$ to the film. Along the P-wall, the left- and right-handed twists alternate, being separated by defect lines along the $z$-direction. As one approaches the tip of the DW, the realignment trajectory of $\mathbf{P}$ starts to resemble a sharp U-turn with a high splay-bend energy; the geometry of the P-wall is replaced by the new structure, called the T-wall, in which the polarization vectors of the two domains are both tangential to it. The T-branch is separated from the P-branches by two -1/2 disclinations. The T-branch is more narrow than the P-branches, which might be explained by a smaller amount of splay in it; the prevailing deformation appears to be the twist along the horizontal and vertical axes with some elements of splay-bend near the bounding plates. Since the T-wall embraces the circular domain on the inside, its eccentricity is less than 1; the curved T-wall causes bending of $\mathbf{P}$ in the outside domain, which produces a "ghost" parabola.

We show that these experimentally observed shapes are well described by a version of the Oseen-Frank energy functional in which the splay elastic constant is much larger than the twist and bend constants. Whenever the third component of the polarization vector is set to vary, the model allows for the presence of both twist- and bend-splay DWs. This approach recovers principal features of ferroelectric nematic morphologies, such as the secondary "ghost" domain walls and mixed handedness of the twist of the polarization vector along the DW. Within the modeling, the finite width of domain walls is attributable to the large disparity between the elastic constants, in that the system accommodates the splay by forming structures with strong bend and/or twist. The finite width of domain walls in the model is fully supported by the experiments, as evident from the comparison of parabolic walls produced by simulations in Figs. 11, 13-17 and their experimental textures in Figs. 1,2,4,5. Similar agreement is evident by comparing the experimental textures of rectilinear domain walls in films freely suspended in square openings, Fig.6, and their simulated structure in Fig.19.

We also note that model (7) allows for bend and splay within the walls, while (8) incorporates all elastic modes, including twists, yet both produce the same DW morphology and director distribution outside of the DWs. At the same time, within the walls, the director configurations are different and are dominated either by bend or by twist. Further, note that in numerical experiments ghost walls seem to appear only in the bend dominated walls, therefore conceivably one could try to use the presence of ghost walls in physical experiments as an indicator of a bend deformation within DWs. Needless to say, the models described in this work are not uniformly applicable in that they fail to capture all the features of ferroelectric nematic configurations when the dependence on the $z$-coordinate cannot be ignored. In particular, this shortcoming applies to the details of the wall structure that in fact might be three-dimensional. The present model also does not capture the expected presence of double layers of bound charges of a density $-\partial P_k/\partial k$ embracing the parabolic DWs, although these are visible around straight DWs in Fig. 9a. The full three-dimensional inner structures of DWs will be addressed in a future investigation.

**Acknowledgements**

This work was supported by the NSF grants DMS-2106551 (DG), DMS-2106516 (PS), and DMR-2341830 (ODL); NASU project No. 0121U109816, the long-term program of support of the Ukrainian research teams at the Polish Academy of Sciences carried out in collaboration with the U.S. National Academy of Sciences with the financial support of external partners via the agreement No. PAN.BFB.S.BWZ.356.022.2023 (VGN), NATO Science for Peace and Security Programme grant G6030 (VGN and ODL), and a Simons Collaboration grant 585520 (PS).

**References**

1. P. S. Bednyakov, B. I. Sturman, T. Sluka, A. K. Tagantsev, P. V. Yudin, Physics and applications of charged domain walls. *Npj Comput Mater* **4**, 65 (2018).
2. G. F. Nataf, M. Guennou, J. M. Gregg, D. Meier, J. Hlinka, E. K. H. Salje, J. Kreisel, Domain-wall engineering and topological defects in ferroelectric and ferroelastic materials. *Nat Rev Phys* **2**, 634-648 (2020).
3. G. Catalan, J. Seidel, R. Ramesh, J. F. Scott, Domain wall nanoelectronics. *Rev Mod Phys* **84**, 119-156 (2012).
4. K. A. Hunnestad, E. D. Roede, A. T. J. van Helvoort, D. Meier, Characterization of ferroelectric domain walls by scanning electron microscopy. *J Appl Phys* **128**, 191102 (2020).
5. L. D. Landau, E. Lifshitz, On the theory of the dispersion of magnetic permeability in ferromagnetic bodies. *Phys. Zeitsch. Sowjetunion* **8**, 153-169 (1935).
6. Anonymous, *Topological Structures in Ferroic Materials: Domain Walls, Vortices and Skyrmions*. R. Hull, Ed., Springer Series in Materials Science (Springer, Switzerland, 2016), vol. 228, pp. 242.
7. R. J. Mandle, S. J. Cowling, J. W. Goodby, A nematic to nematic transformation exhibited by a rod-like liquid crystal. *Phys Chem Chem Phys* **19**, 11429-11435 (2017).
8. H. Nishikawa, K. Shiroshita, H. Higuchi, Y. Okumura, Y. Haseba, S.-I. Yamamoto, K. Sago, H. Kikuchi, A Fluid Liquid-Crystal Material with Highly Polar Order. *Adv Mater* **29**, 1702354 (2017).
9. N. Sebastián, L. Cmok, R. J. Mandle, M. R. de la Fuente, I. D. Olenik, M. Čopič, A. Mertelj, Ferroelectric-Ferroelastic Phase Transition in a Nematic Liquid Crystal. *Phys Rev Lett* **124**, 037801 (2020).
10. X. Chen, E. Korblova, D. P. Dong, X. Y. Wei, R. F. Shao, L. Radzihovsky, M. A. Glaser, J. E. Maclennan, D. Bedrov, D. M. Walba, N. A. Clark, First-principles experimental demonstration of ferroelectricity in a thermotropic nematic liquid crystal: Polar domains and striking electro-optics. *Proceedings of the National Academy of Sciences of the United States of America* **117**, 14021-14031 (2020).
11. P. G. de Gennes, J. Prost, *The Physics of Liquid Crystals* (Clarendon Press, Oxford, 1993), pp. 598.
12. X. Chen, E. Korblova, M. A. Glaser, J. E. Maclennan, D. M. Walba, N. A. Clark, Polar in-plane surface orientation of a ferroelectric nematic liquid crystal: Polar monodomains and twisted state electro-optics. *Proceedings of the National Academy of Sciences of the United States of America* **118**, e2104092118 (2021).
13. J. X. Li, H. Nishikawa, J. Kougo, J. C. Zhou, S. Q. Dai, W. T. Tang, X. H. Zhao, Y. Hisai, M. J. Huang, S. Aya, Development of ferroelectric nematic fluids with giant-epsilon dielectricity and nonlinear optical properties. *Science Advances* **7**, eabf5047 (2021).
14. P. Rudquist, Revealing the polar nature of a ferroelectric nematic by means of circular alignment. *Sci Rep-UK* **11**, 24411 (2021).

15. F. Caimi, G. Nava, R. Barboza, N. A. Clark, E. Korblova, D. M. Walba, T. Bellini, L. Lucchetti, Surface alignment of ferroelectric nematic liquid crystals. *Soft Matter* **17**, 8130-8139 (2021).
16. S. Brown, E. Cruickshank, J. M. D. Storey, C. T. Imrie, D. Pociecha, M. Majewska, A. Makal, E. Górecka, Multiple Polar and Non-polar Nematic Phases. *Chemphyschem* **22**, 2506-2510 (2021).
17. N. Sebastián, R. J. Mandle, A. Petelin, A. Eremin, A. Mertelj, Electrooptics of mm-scale polar domains in the ferroelectric nematic phase. *Liq Cryst* **48**, 2055-2071 (2021).
18. X. Chen, Z. C. Zhu, M. J. Magrini, E. Korblova, C. S. Park, M. A. Glaser, J. E. Maclennan, D. M. Walba, N. A. Clark, Ideal mixing of paraelectric and ferroelectric nematic phases in liquid crystals of distinct molecular species. *Liq Cryst* **49**, 1531-1544 (2022).
19. B. Basnet, M. Rajabi, H. Wang, P. Kumari, K. Thapa, S. Paul, M. O. Lavrentovich, O. D. Lavrentovich, Soliton walls paired by polar surface interactions in a ferroelectric nematic liquid crystal. *Nature Communications* **13**, 3932 (2022).
20. N. Sebastián, M. Čopič, A. Mertelj, Ferroelectric nematic liquid-crystalline phases. *Phys Rev E* **106**, 021001 (2022).
21. X. Chen, V. Martinez, P. Nacke, E. Korblova, A. Manabe, M. Klasen-Memmer, G. Freychet, M. Zhernenkov, M. A. Glaser, L. Radzihovsky, J. E. Maclennan, D. M. Walba, M. Bremer, F. Giesselmann, N. A. Clark, Observation of a uniaxial ferroelectric smectic A phase. *Proceedings of the National Academy of Sciences of the United States of America* **119**, e2210062119 (2022).
22. P. Kumari, B. Basnet, H. Wang, O. D. Lavrentovich, Ferroelectric nematic liquids with conics. *Nature Communications* **14**, 748 (2023).
23. D. R. Link, N. Chattham, J. E. Maclennan, N. A. Clark, Effect of high spontaneous polarization on defect structures and orientational dynamics of tilted chiral smectic freely suspended films. *Phys Rev E* **71**, 021704 (2005).
24. O. D. Lavrentovich, Splay-bend elastic inequalities shape tactoids, toroids, umbilics, and conic section walls in paraelectric, twist-bend, and ferroelectric nematics. *Liquid Crystals Reviews* **12**, 1-13 (2024).
25. Apollonius of Perga, T. L. Heath, *Treatise on Conic Sections* (Cambridge University Press, 1896).
26. Z. Zhuang, J. E. Maclennan, N. A. Clark, Device Applications of Ferroelectric Liquid-Crystals - Importance of Polarization Charge Interactions. *P Soc Photo-Opt Ins* **1080**, 110-114 (1989).
27. A. Pattanaporkratana (2008) Textures and interactions between vortices in the 2D XY field of freely suspended SmC and SmC* liquid crystal films, PhD thesis. Physics Department (University of Colorado, Boulder, CO), p 133.
28. S. Cherifi-Hertel, C. Voulot, U. Acevedo-Salas, Y. D. Zhang, O. Crégut, K. D. Dorkenoo, R. Hertel, Shedding light on non-Ising polar domain walls: Insight from second harmonic generation microscopy and polarimetry analysis. *J Appl Phys* **129**, 081101 (2021).
29. R. B. Meyer, L. Liebert, L. Strzelecki, P. Keller, Ferroelectric Liquid-Crystals. *J Phys Lett-Paris* **36**, L69-L71 (1975).
30. R. B. Meyer, Ferroelectric Liquid-Crystals - Review. *Mol Cryst Liq Cryst* **40**, 33-48 (1977).

31. K. Okano, Electrostatic Contribution to the Distortion Free-Energy Density of Ferroelectric Liquid-Crystals. *Jpn J Appl Phys 2* **25**, L846-L847 (1986).
32. J. B. Lee, R. A. Pelcovits, R. B. Meyer, Role of electrostatics in the texture of islands in free-standing ferroelectric liquid crystal films. *Phys Rev E* **75**, 051701 (2007).
33. X. Chen, V. Martinez, E. Korblova, G. Freychet, M. Zhernenkov, M. A. Glaser, C. Wang, C. H. Zhu, L. Radzihovsky, J. E. Maclennan, D. M. Walba, N. A. Clark, The smectic ZA phase: Antiferroelectric smectic order as a prelude to the ferroelectric nematic. *Proceedings of the National Academy of Sciences of the United States of America* **120**, e2217150120 (2023).
34. A. Adaka, M. Rajabi, N. Haputhantrige, S. Sprunt, O. D. Lavrentovich, A. Jákli, Dielectric Properties of a Ferroelectric Nematic Material: Quantitative Test of the Polarization-Capacitance Goldstone Mode. *Phys Rev Lett* **133**, 038101 (2024).
35. A. Erkoreka, J. Martinez-Perdiguero, Constraining the value of the dielectric constant of the ferroelectric nematic phase. *Phys Rev E* **110**, L022701 (2024).
36. N. A. Clark, X. Chen, J. E. MacLennan, M. A. Glaser, Dielectric spectroscopy of ferroelectric nematic liquid crystals: Measuring the capacitance of insulating interfacial layers. *Physical Review Research* **6**, 013195 (2024).
37. A. Mertelj, L. Cmok, N. Sebastián, R. J. Mandle, R. R. Parker, A. C. Whitwood, J. W. Goodby, M. Čopič, Splay Nematic Phase. *Phys Rev X* **8**, 041025 (2018).
38. O. O. Ramdane, P. Auroy, S. Forget, E. Raspaud, P. Martinot-Lagarde, I. Dozov, Memory-free conic anchoring of liquid crystals on a solid substrate. *Phys Rev Lett* **84**, 3871-3874 (2000).
39. R. Oldenbourg, A new view on polarization microscopy. *Nature* **381**, 811-812 (1996).
40. M. Shribak, R. Oldenbourg, Techniques for fast and sensitive measurements of two-dimensional birefringence distributions. *Appl Optics* **42**, 3009-3017 (2003).
41. O. D. Lavrentovich, Looking at the world through liquid crystal glasses. *Contemporary Mathematics* **577**, 25-46 (2012).
42. Y. K. Kim, G. Cukrov, J. Xiang, S. T. Shin, O. D. Lavrentovich, Domain walls and anchoring transitions mimicking nematic biaxiality in the oxadiazole bent-core liquid crystal C7. *Soft Matter* **11**, 3963-3970 (2015).
43. O. D. Lavrentovich, Geometrical Anchoring at an Inclined Surface of a Liquid-Crystal. *Phys Rev A* **46**, R722-R725 (1992).
44. P. Kumari, B. Basnet, M. O. Lavrentovich, O. D. Lavrentovich, Chiral ground states of ferroelectric liquid crystals. *Science* **383**, 1364-1368 (2024).
45. S. Yi, Z. Hong, Z. Ma, C. Zhou, M. Jiang, X. Huang, M. Huang, S. Aya, R. Zhang, Q.-H. Wei, Chiral pi domain walls composed of twin half-integer surface disclinations in ferroelectric nematic liquid crystals. *ArXiv*, arXiv:2406.13326v13321 (2024).
46. M. Kléman, *Points, lines and walls in liquid crystals, magnetic systems and various ordered media* (John Wiley & Sons, Chichester, 1983), pp. 322.
47. R. N. Thurston, J. Cheng, R. B. Meyer, G. D. Boyd, Physical-Mechanisms of Dc Switching in a Liquid-Crystal Bistable Boundary-Layer Display. *J Appl Phys* **56**, 263-272 (1984).
48. V. G. Nazarenko, O. D. Lavrentovich, Anchoring Transition in a Nematic Liquid Crystal Composed of Centrosymmetric Molecules. *Phys Rev E* **49**, R990-R993 (1994).

49. F. Caimi, G. Nava, S. Fuschetto, L. Lucchetti, P. Paie, R. Osellame, X. Chen, N. A. Clark, M. A. Glaser, T. Bellini, Fluid superscreening and polarization following in confined ferroelectric nematics. *Nat Phys* **19**, 1658-1666 (2023).
50. D. Golovaty, P. Sternberg, R. Venkatraman, A Ginzburg-Landau-Type Problem for Highly Anisotropic Nematic Liquid Crystals. *Siam J Math Anal* **51**, 276-320 (2019).
51. F. John, *Partial differential equations*, Applied Mathematical Sciences (Springer-Verlag, New York, ed. 4, 1991), vol. 1.
52. COMSOL Multiphysics v.5.3; http://www.comsol.com/, Stockholm, Sweden.